\RequirePackage{fix-cm}
\documentclass[twocolumn]{svjour3}          
\smartqed  
\usepackage{graphicx}
\usepackage{amsmath}
\usepackage{amssymb, amsfonts}
\usepackage{algorithm,algorithmic}
\usepackage{psfrag,epsf}
\usepackage{graphics}
\usepackage{enumerate}
\usepackage{booktabs}
\usepackage{multirow}
\usepackage{color}
\usepackage{bbm}
\usepackage{parskip}
\usepackage{tikz}
\usepackage{geometry}
\usepackage[export]{adjustbox}
\PassOptionsToPackage{hyphens}{url}\usepackage[colorlinks=true,citecolor=blue,urlcolor=violet]{hyperref}
\usepackage{caption}
\usepackage{subcaption}
\usepackage{mathtools}
\usepackage[toc,page]{appendix}
\usepackage{verbatim}
\usepackage{pdfpages}
\usepackage{lscape}
\usepackage{lineno}
\usepackage{setspace}
\usepackage{array}
\usepackage{natbib}
\newcommand{\btheta}{ \boldsymbol\theta} 
\newcommand{\bbeta}{ \boldsymbol\beta} 
\newcommand{\bpsi}{\boldsymbol\psi} 
\newcommand{\bepsilon}{ \boldsymbol\epsilon} 
\newcommand{\bDelta}{ \boldsymbol\Delta } 
\newcommand{\bSigma}{\boldsymbol{\Sigma} }
\newcommand{\bGamma}{ \boldsymbol\Gamma} 
\newcommand{\bTheta}{ \boldsymbol\Theta} 

\newcommand{\bA}{ \mathbf{A}}

\newcommand{\be}{ \mathbf{e}}

\newcommand{\bx}{ \mathbf{x}}
\newcommand{\bX}{ \mathbf{X}}
\newcommand{\bB}{ \mathbf{B}}
\newcommand{\bZ}{ \mathbf{Z}}
\newcommand{\by}{ \mathbf{y}}

\newcommand{\bz}{ \mathbf{z}}

\newcommand{\br}{ \mathbf{r}}

\newcommand{\bs}{ \mathbf{s}}
\newcommand{\bb}{ \mathbf{b}}

\newcommand{\bF}{ \mathbf{F}}
\newcommand{\bu}{ \mathbf{u}}
\newcommand{\bU}{ \mathbf{U}}
\newcommand{\bv}{ \mathbf{v}}
\newcommand{\bV}{ \mathbf{V}}

\newcommand{\bI}{ \mathbf{I}}
\newcommand{\bD}{ \mathbf{D}}
\newcommand{\bM}{\mathbf{M}} 
\newcommand{\bW}{ \mathbf{W}}
\newcommand{\bw}{ \mathbf{w}}

\newcommand{\bR}{ \mathbf{R}}

\newcommand{\calL}{\mathcal{L}}

\newcommand{\calS}{\mathcal{S}}

\newcommand{\calK}{\mathcal{K}}

\newcommand{\Matern}{ \mbox{Mat$\acute{\mbox{e}}$rn}}

\newcommand{\beq}{ \begin{equation}}
\newcommand{\eeq}{ \end{equation}}
\newcommand{\beqn}{ \begin{eqnarray}}
\newcommand{\eeqn}{ \end{eqnarray}}
\newcommand{\Lp}{\left(}
\newcommand{\Rp}{\right)}
\newcommand{\lp}{\left[}
\newcommand{\rp}{\right]}

\newcommand{\rone}{\mathbb{R}}
\newcommand{\rtwo}{\mathbb{R}^2}
\newcommand{\rd}{\mathbb{R}^d}
\newcommand{\rn}{\mathbb{R}^n}

\newcommand{\logdet}{\log\det\,}
\newcommand{\logdethat}{\widetilde{\log\det}\,}
\journalname{Statistics and Computing}
\begin{document}

\title{Kryging: Geostatistical analysis of large-scale datasets using Krylov subspace methods}

\titlerunning{Kryging}        

\author{Suman Majumder         \and
        Yawen Guan              \and
        Brian J. Reich          \and
        Arvind K. Saibaba
}

\authorrunning{Majumder et al.} 

\institute{Suman Majumder \at
              Department of Statistics, North Carolina State University \\
              \email{smajumder@hsph.harvard.edu}             \\
             \emph{Present address:} Department of Nutrition, Harvard T.H. Chan School of Public Health
           \and
           Yawen Guan \at
              Department of Statistics, University of Nebraska-Lincoln\\
              \email{yguan12@unl.edu}
              \and
              Brian J. Reich \at
              Department of Statistics, North Carolina State University \\
              \email{bjreich@ncsu.edu}
              \and
              Arvind K. Saibaba \at
              Department of Mathematics, North Carolina State University \\
              \email{asaibab@ncsu.edu}
}

\date{Received: date / Accepted: date}

\maketitle

\begin{abstract}
Analyzing massive spatial datasets using a Gaussian process model poses computational challenges. This is a problem prevailing heavily in applications such as environmental modeling, ecology, forestry and environmental health. We present a novel approximate inference methodology that uses profile likelihood and Krylov subspace methods to estimate the spatial covariance parameters and makes spatial predictions with uncertainty quantification for point-referenced spatial data. The proposed method, Kryging, applies for both observations on regular grid and irregularly-spaced observations, and for any Gaussian process with a stationary isotropic (and certain geometrically anisotropic) covariance function, including the popular $\Matern$ covariance family. We make use of the block Toeplitz structure with Toeplitz blocks of the covariance matrix and use fast Fourier transform methods to bypass the computational and memory bottlenecks of approximating log-determinant and matrix-vector products. We perform extensive simulation studies to show the effectiveness of our model by varying sample sizes, spatial parameter values and sampling designs. A real data application is also performed on a dataset consisting of land surface temperature readings taken by the MODIS satellite. Compared to existing methods, the proposed method performs satisfactorily with much less computation time and better scalability.
\keywords{Approximate inference \and Profile likelihood \and Block Toeplitz matrix \and Fast Fourier transform \and Krylov subspace methods \and Golub-Kahan bidiagonalization}
\end{abstract}

\section{Introduction}
\label{sec:intro}


Massive spatial datasets, often coming from satellites or other remotely-sensed sources, have become increasingly common in applications such as environmental health, forestry, ecology etc. Classical geostatistical analysis methods for point-referenced spatial data are burdened with computationally intensive steps such as Cholesky factorization or eigendecomposition which have cubic complexity in the number of observations. Despite the advances in computing performance, these methods remain prohibitively expensive to apply to datasets of even moderately-large size. Therefore, we need to develop methods that perform nearly as well as the classical methods but are more computationally efficient and therefore applicable to problems of massive volume.

There is a rich literature of approximate inference methods for point-referenced spatial data. Early approaches approximated the joint likelihood by decomposing it into a product of conditional distributions \citep{vecchia1988estimation, stein2004approximating}, using pseudo-likelihood \citep{varin2011overview, eidsvik2014estimation} or using covariance tapering \citep{furrer2006covariance, kaufman2008covariance, stein2013statistical}. Modeling in the spectral domain \citep{fuentes2007approximate, guinness2017circulant, guinness2019spectral} was also used to circumvent the heavy computation. Another class of approaches are based on finite-rank approximations such as fixed-rank Kriging  \citep{cressie2008fixed, kang2011bayesian, katzfuss2011spatio}, predictive process \citep{banerjee2008gaussian, finley2009improving}, process convolution \citep{higdon2002space} and lattice Kriging \citep{nychka2015multiresolution}. Other approaches use a combination of hierarchical matrix approaches and stochastic estimators for the log-likelihood~\citep{anitescu2012matrix, ambikasaran2015fast,minden2017fast,eriksson2018scaling,stein2013statistical} or spectral methods and h-likelihood \citep{dutta2016reml}.

More recent approaches make use of the modern computing platforms and focus on parallelizing the computational load. \cite{paciorek2015parallelizing} is one such example. \cite{katzfuss2017multi} and \cite{katzfuss2017parallel} combine low rank methods with distributed computing. Dividing the data into subsets, drawing inference on these subsets in parallel and recombining them has been proposed by \cite{barbian2017spatial} and \cite{guhaniyogi2018meta}. \cite{datta2016hierarchical,datta2016nearest,datta2016nonseparable} use an approximation based on the conditional distribution given the nearest neighbors, inducing sparsity and allowing the method to be parallelized. The stochastic partial differential equation or SPDE \citep{lindgren2011explicit} approach induces sparsity in the inverse-covariance matrix for fast approximations. \cite{sun2012geostatistics}, \cite{bradley2016comparison}, \cite{heaton2019case} and \cite{liu2020gaussian} provide comprehensive reviews of these methods and demonstrate their effectiveness in spatial modeling.

Most of these methods use either finite-rank approximations or introduce sparsity in the covariance or the inverse-covariance structure. Finite rank based models typically have complexity $\mathcal{O}(nr^2 + r^3)$ with $r$ being the rank of the model such that $r \ll n$. However, in order for the approximation to be effective for large $n$, a large rank $r$ is needed which increases the computational costs. This cost can be alleviated by inducing sparsity into the covariance structure using compactly supported covariance function; however, this may not be an appropriate modeling choice when long-range dependence is present in the data.

 We present a novel statistical method of log-linear complexity to provide approximate inference for massive geostatistical datasets using profile maximum likelihood estimation and Krylov subspace methods based on the genHyBR method proposed by \cite{chung2018efficient}. The proposed method, Kryging, provides prediction for the observed process at unobserved locations by approximating the underlying spatial process on a regular, equispaced grid. Although we approximate the latent process on a grid, we do not restrict the observations to be on grid and therefore the method can be applied to irregularly-spaced large spatial datasets. We generate estimates of the underlying process through Krylov subspace methods. Krylov subspaces \citep[See ][ for reference]{MR1990645} are efficient iterative methods for solving large-scale linear systems and least squares problems. A key advantage of the Krylov subspace approach is that it is matrix-free, in that it does not require forming the matrices explicitly, but only requires the action of the matrix on appropriate vectors. We provide prediction uncertainty estimates in the form of pointwise $95\%$ confidence intervals via a parametric bootstrap approach and estimates for the mean and spatial covariance parameters. Kryging applies to any stationary isotropic covariance structure, e.g., the Mat\'ern covariance family, as well as covariance functions that incorporate geometric anisotropy by allowing dissimilar stretching along the two axes. It exploits the Toeplitz (in one dimension) or block Toeplitz with Toeplitz blocks (BTTB) structure (in higher dimensions) of the resulting covariance matrices and employs a fast Fourier transformation based method for achieving computational gains for matrix-vector multiplications \citep[See ][]{gray2006toeplitz} and approximating log-determinants \citep{kent1996spectral}. As a result, Kryging has $\mathcal{O}(n)$ storage costs and only $\mathcal{O}(n \log n)$ computational complexity where $n$ is the size of the underlying grid for estimating the spatial parameters and performing spatial prediction.

\textcolor{black}{The tools used for building the Kryging model have been used in literature before in different contexts and different problems. However, by efficiently combining them in a specific manner, Kryging has several advantages compared to related methods in the literature.} \cite{chung2018efficient} also use the same core method but we extend it to include mean and spatial covariance parameter estimation, uncertainty quantification and approximation of log-determinants. \cite{aune2014parameter} and \cite{dutta2016reml}  also use tools such as Krylov subspaces and the fast Fourier transformation, but their usage differs vastly from ours. \textcolor{black}{First, we construct a different Krylov subspace, one that incorporates the noise covariance, a mapping matrix, and the covariance matrix; in contrast, the approach in the other papers is to build a Krylov subspace method with the covariance matrix alone. Second, we use the Golub-Kahan bidiagonalization rather than Lanczos or Conjugate Gradient for linear systems. Third, we use the basis vectors from the Krylov subspace to estimate the objective function and the gradients (one exception is the determinant and its derivative for which we use a different approximation). In contrast, other approaches use various tools such as Monte Carlo trace estimators, to estimate the various quantities.}

Kryging has a low-rank matrix involved in the approximation process. However, compared to other low-rank methods discussed above, empirical evidence hints that using a small order of the Krylov subspace works well for huge datasets and produces accurate results. \textcolor{black}{Block-circulant embeddings has been proposed as a stand-alone method to approximate determinants \citep{rue2005gaussian} which nicely gels with the Krylov subspace based approximation to the problem of maximizing the quadratic part of a Gaussian log-likelihood to produce a fast and scalabe approximate inference method for massive geostatistical datasets.}

 We establish the particular form of latent Gaussian model that we use for our method in Section \ref{sec:obs_mod}. Section \ref{sec: infer} gives the details of the method. We provide detailed description and algorithms of components of the method in various subsections of Section~\ref{sec: infer}. A thorough simulation study is performed in Section \ref{sec:sim} and an application to MODIS satellite data is performed in Section \ref{sec:apply}. The data analysis is based on \cite{heaton2019case}. The rationale behind this was to be able to compare the performance of our method to other available methods directly. We finish with a discussion and concluding remarks in Section \ref{sec:conc}.

\section{Latent Gaussian Model}
\label{sec:obs_mod}


Let $y(\bs)$ be the observed process and $x(\bs)$ is the underlying process of interest at location $\bs \in \rd$, $d \geq 1$\textcolor{black}{; throughout this paper, we illustrate the methods using the $d=2$ but our approach is applicable to problems with two or three spatial dimensions with a possible additional time dimension.} A realization from the observation process, $\by =$ $[y(\bs_1), \ldots, y(\bs_p)]^{\sf T}$, at $p$ locations $\bs_1, \ldots, \bs_p$ is related to a realization from the latent process, $\bx = [x(\bs^*_1), \ldots, x(\bs^*_n)]^{\sf T}$, at $n$ possibly different locations $\bs^*_1, \ldots, \bs^*_n$ by the relationship \beq \label{eq: obs_mod} \by = \bX\bbeta + \bA\bx + \bepsilon, \eeq where $\bepsilon \sim N\Lp \mathbf{0}, \bR\Rp$ with $\bR$ being a cheaply invertible matrix of individual variances for each location, $\bX$ being the matrix of corresponding covariates observed at the same locations as the observations themselves and $\bA$ being a matrix that specifies the linear combinations that connect the mean removed $\by$ and $\bx$. \textcolor{black}{For this paper, we make the standard assumption that the nugget variance is constant across space and set $\mathbf{R}=\tau^2\mathbf{I}_p$.}

The mapping matrix $\bA$ permits the flexibility of $\by$ and $\bx$ not being co-located, as well as change of support. For example, $\bA = \bI$, the identity matrix, represents the case where $\by$ is a noisy observation of $\bx$ itself after accounting for the mean process. In case the response locations are a subset of the $n$ locations $\bs^*_1, \ldots, \bs_n^*$, then $\bA$ is the $n \times n$ identity matrix with $n-p$ rows removed. The matrix $\bA$ can be non-diagonal as well, for the case when value of $\by$ at each location is considered as an average of the unobserved $\bx$ at nearby locations, as it can be when $y(\bs)$ is observed at locations at irregularly spaced locations and $x(\bs)$ is considered on a grid around those locations. 

 When the observations are not on a regular grid, we still set the latent process locations $\bs_1^*, \ldots, \bs_n^*$ to be on a rectangular grid and account for the irregularity of the observation locations in the mapping matrix, $\bA$.  We specify the entries of $\bA$ so that each observation is a convex combination of the latent process in the neighborhood of the observation.  Specifically, the latent process is weighted by the Wendland kernel function \citep{wendland1995piecewise} $w(d_{ij}) = (1-d_{ij})^4_+\Lp 1 + 4d_{ij} \Rp$, where $d_{ij} = \text{ max }\{ |s_{i1} - s_{j1}^*|/\Delta_1, |s_{i2} - s_{j2}^*|/\Delta_2 \}$ and $(x)_+ = \max\{x,0\}$, $\Delta_1$ and $\Delta_2$ are the grid spacings in the two directions and $\bs_i=(s_{i1},s_{i2})$ and $\bs_j^*=(s_{j1}^*,s_{j2}^*)$ are the $i$-th observation location and $j$-th grid-point location, respectively. This particular formulation allows to approximate the value at a point outside of the grid as a weighted combination of its nearest four neighbors while for a point on the grid itself, the approximation is exact. To ensure the weights are convex, they are normalized to sum to one for each observation.  That is, we assume the mean response is 
$$E\{y(\bs_i)\} = X(\bs_i)^{\sf T}\bbeta + \frac{\sum_{j=1}^nw(d_{ij})x(\bs_j^*)}{\sum_{k=1}^nw(d_{ik})}.$$
This is equivalent to setting the $(i,j)$ element of $\bA$ to $w(d_{ij})/\{\sum_{k=1}^nw(d_{ik})\}$. The truncation function $(x)_+$ ensures that $\bA$ is a sparse matrix with at most four nonzero entries per row, i.e., the matrix $\bA$ has $\mathcal{O}(p)$ nonzero entries. 

Choosing the mapping matrix to be sparse ensures there is not significantly higher computational cost due to these changes when applying to an irregularly spaced data. This approach to handling irregularly-spaced observations introduces an additional tuning parameter, $n$, which controls the density of the latent space observations.  When the observation locations are on a regular grid, we simply set it to be equal to  $p$ so that the latent process locations match the observations.  However, when the observations are not on a grid then there is no natural choice for $n$. Accuracy should increase with $n$ at the expense of computational burden.  This issue is explored further in the simulation study of Section \ref{sec:sim}.

We use a latent Gaussian process to model the true state $\bx(\bs)$, with zero mean and isotropic $\Matern$ covariance kernel \citep{matern1960spatial} with standard deviation $\sigma$, spatial range parameter $\rho$ and  smoothness parameter $\nu$. Therefore, at finite collection of locations, $\bx$ is a multivariate Gaussian distribution with mean $\mathbf{0}$ and $n \times n$ correlation matrix $\bSigma$, i.e., \beq \bx \sim N\Lp \mathbf{0}, \sigma^2\bSigma \Rp,\eeq with $\mathbf{0}$ being the vector of all zeros and \[\bSigma_{ij} = \frac{2^{1-\nu}}{\Gamma(\nu)}\Lp \sqrt{2\nu}\frac{d_{ij}}{\rho} \Rp^{\nu} K_{\nu} \Lp \sqrt{2\nu}\frac{d_{ij}}{\rho} \Rp\] being the spatial correlation between locations $i$ and $j$ induced by the stationary isotropic $\Matern$ covariance kernel for $i, j = 1, \ldots, n$. Here $d_{ij} = \|\bs_i - \bs_j\|_2$ and $\| \cdot \|_2$ denotes the Euclidean norm in $\rtwo$ and $K_\nu(\cdot)$ is the modified Bessel function of the second kind with parameter $\nu$. The choice of $\Matern$ covariance kernel is common but any other stationary covariance function (or geometrically anisotropic covariance function that induces different stretching along the two axes) may be used along with the approach for both regularly gridded and irregularly spaced datasets with same computational complexity that we outline in the next section.

\section{Inferential Approach}
\label{sec: infer}


In this section, we describe an inferential approach for the latent Gaussian model that combines Kriging and Krylov subspace methods, which we have been calling ``Kryging''. The likelihood function for the latent state $\bx$ and the mean and spatial variance parameters $\btheta = \Lp \bbeta^{\sf T}, \sigma^2, \tau^2, \rho, \nu \Rp^{\sf T}$  can be written as\beq \calL(\bx, \btheta; \by) = f_{\by,\btheta}(\by|\bx)f_{\bx,\btheta}(\bx|\btheta),\eeq where $f_{\by,\btheta}(\cdot|\bx)$ is the density of the data given $\bx$ and $f_{\bx,\btheta}(\cdot)$ is the density of $\bx$; both densities depend on the parameter $\btheta$.  Since we assumed a Gaussian model for $\by|\bx$ and $\bx$, we have 
\[\log f_{\by,\btheta}(\by|\bx) \simeq -\frac{p}{2}\log \tau^2 -\frac{1}{2\tau^2}\bpsi^{\sf T}\bpsi,
\] where $\simeq$ means equal up to a constant that is unimportant for the purposes of optimization and $\bpsi = \by - \bX\bbeta - \bA\bx$  and\beq \begin{split}
    \log f_{\bx,\btheta}(\bx) & \simeq -\frac{n}{2} \log \sigma^2 -\frac{1}{2}\logdet(\bSigma(\btheta))\\
    & \quad -\frac{1}{2\sigma^2}\bx^{\sf T}\bSigma(\btheta)^{-1} \bx.
\end{split}\eeq Thus the log-likelihood function, $l(\bx, \btheta) =\log $ $\calL(\bx, \btheta; \by)$ has the form \beq \label{eq: exact_lik} \begin{split}
    l(\bx,\btheta) & \simeq -\frac{p}{2}\log \tau^2 -\frac{1}{2\tau^2}\bpsi^{\sf T}\bpsi\\
    & \quad-\frac{n}{2}\log \sigma^2 - \frac{1}{2}\logdet\bSigma(\btheta)\\
    & \quad - \frac{1}{2\sigma^2}\bx^{\sf T}\bSigma(\btheta)^{-1}\bx.
\end{split}  \eeq

Evaluation of the log-likelihood function involves inverting and computing the log-determinant of the covariance matrix $\bSigma(\btheta)$, both of which require $\mathcal{O}(n^3)$ many operations which is not feasible for large $n$. Since the optimization needs to run on both $\bx$ and $\btheta$, it would be a ultra high-dimensional optimization which would generally be infeasible to implement. Therefore, running an optimization procedure over both $\btheta$ and $\bx$ on this objective function straightaway is futile and we must look into approximation methods to avoid these computational bottlenecks.

We propose a computationally-efficient inference approach using approximate inference for fast estimation for both parameters $\btheta$ and the underlying true state variables $\bx$ along with its uncertainty. We profile $\bx$ as a function of the parameters $\btheta$ and maximize the corresponding profile likelihood over $\btheta$ \citep{cox1989analysis}. This reduces the dimensionality of the optimization problem greatly but it requires an estimate of $\bx$ for a given value of $\btheta$. 

The genHyBR method \citep{chung2018efficient} circumvents the matrix inversion problem as it brings down the total complexity of computing the quadratic term to that of a matrix vector multiplication. Typically this would take $\mathcal{O}(n^2)$ operations. However, computational techniques such as Fast Fourier Transforms (FFTs) or $\mathcal{H}$-matrices (a review of techniques can be found in~\cite{ambikasaran2015fast}) can reduce the computational cost of storage and the mathematical operators to $\mathcal{O}(n\log^r n)$, where $r$ is a non-negative exponent which depends on the operation and the method used. In particular, we use the symmetric BTTB structure of $\bSigma(\btheta)$. The symmetric BTTB structure allows us to store $\bSigma(\btheta)$ in $\mathcal{O}(n)$, since only one row/column of $\bSigma(\btheta)$ needs to be stored, and  compute the matrix vector products involving $\bSigma(\btheta)$ in $\mathcal{O}(n \log n)$ time. If the underlying process realizations are not on a regular grid, then the $\mathcal{H}$-matrix approach can be used instead with the same computational cost. However, with the mapping matrix strategy laid out in Section \ref{sec:obs_mod}, we do not require this approach. The symmetric BTTB structure also allows us to compute the log-determinant of $\bSigma(\btheta)$ in $\mathcal{O}(n \log n)$ time. This gives us a good estimate for $\bx$ for a given value of $\btheta$.

\subsection{Profile Likelihood}
\label{subsec:prof_lik}

Maximizing the log-likelihood function in Eq. (\ref{eq: exact_lik}) as a function of both $\bx$ and $\btheta$ is not feasible and therefore we use a profile likelihood based optimization strategy by profiling $\bx$ as a function of $\btheta$. Profiling out $\bx$ from Eq. (\ref{eq: exact_lik}) as a function of $\btheta$, in exact arithmetic, results in \beq \label{eq: act_xhat}
\begin{split}
    \widehat{\bx}(\btheta) &=\Lp \frac{1}{\sigma^2}\bSigma(\btheta)^{-1} + \frac{1}{\tau^2}\bA^{\sf T}\bA \Rp^{-1}\\
    & \qquad \Lp \frac{1}{\tau^2}\bA^{\sf T}(\by - \bX\bbeta) \Rp .
\end{split} \eeq  
Plugging in $\widehat{\bx}(\btheta)$ in Eq. (\ref{eq: exact_lik}) and calling $\widehat{\bpsi}(\btheta) = \by - \bX\bbeta - \bA\widehat{\bx}(\btheta)$ produces the exact profile log-likelihood function \beq \label{eq: exact_plik}  \begin{split}
    \text{pl}(\btheta) & \simeq -\frac{p}{2}\log \tau^2 -\frac{1}{2\tau^2}\widehat{\bpsi}(\btheta)^{\sf T}\widehat{\bpsi}(\btheta)\\
    & \quad-\frac{n}{2}\log \sigma^2 - \frac{1}{2}\logdet\bSigma(\btheta) -\\ 
    & \quad \frac{1}{2\sigma^2}\widehat{\bx}(\btheta)^{\sf T}\bSigma(\btheta)^{-1}\widehat{\bx}(\btheta).
\end{split} \eeq Since simply evaluating this function involves computing inverses and determinants of the dense covariance matrix, it must be approximated.

Evaluating the exact profile likelihood presents three computational challenges: (1) computing $\widehat{\bx}(\btheta)$ involves inverting large dense $n \times n$ matrices, (2) computing the quadratic term $\widehat{\bx}(\btheta)^{\sf T}\bSigma(\btheta)^{-1}\widehat{\bx}(\btheta)$ and (3) computing the log-determinant of $\bSigma(\btheta)$. The first two are overcome using the genHyBR method \citep{chung2018efficient} while the log-determinant term is approximated using the symmetric BTTB structure of the resulting covariance matrix from the choice of appropriate covariance function previously mentioned in Section \ref{sec:obs_mod}. Once these approximations are in place, optimization of an approximated profile likelihood function can be performed using typical optimization routines to get the estimates of $\btheta$ and $\bx$.

\subsection{genHyBR Method}
\label{subsec:genHyBR}

A key component in maximizing the profile likelihood is to quickly compute $\widehat{\bx}(\btheta) = \underset{\bx}{\text{argmin }}$ $l(\bx,\btheta)$ for a given $\btheta$. The computation of $\widehat{\bx}(\btheta)$ in this context is tantamount to computing \beq \label{eq: genHyBR_eq} \widehat{\bx}(\btheta) = \underset{\bx \in \rn}{\text{ argmin }} \frac{1}{\tau^2}\|\bpsi\|^2_{2} + \frac{1}{\sigma^2}\|\bx\|^2_{\bSigma(\btheta)^{-1}}, \eeq where $\|\br\|^2_{\bM} = \br^{\sf T}\bM \br$ and $\|\cdot\|_2$ represents the Euclidean norm. The genHyBR algorithm \citep{chung2018efficient} solves this weighted least squares problem iteratively using  generalized Golub-Kahan bidiagonalization which is a special type of Krylov subspace method~\citep{benbow1999solving,chung2017generalized}. To simplify notation, we drop the dependence on $\btheta$ and write $\bSigma =\bSigma(\btheta)$.

We provide an outline of the algorithm here. Denote $\calK_k(\bM,\br) =  \text{ span} \{ \br, \bM\br, \ldots, \bM^{k-1}\br \}$ as the Krylov subspace of degree $k$. Observing that Eq.~\eqref{eq: genHyBR_eq} involves the inverse of $\bSigma$, employing a change of variables $\bw = \bSigma^{-1}\bx$ and $\bb = \by - \bX\bbeta$, we then compute $\widehat{\bx}(\btheta) = \bSigma \widehat{\bw}(\btheta)$ and \beq \label{eq: eqv_opt1} \widehat{\bw}(\btheta) = \underset{\bw \in \rn}{\text{ argmin }} \frac{1}{\tau^2}\|\bA\bSigma\bw - \bb\|_2^2 + \frac{1}{\sigma^2}\|\bw\|^2_{\bSigma} . \eeq

Then, for our problem of estimating $\bx$, the genHyBR method \citep{chung2018efficient} looks for the solution of $\bw$ in $$\calS_k = \calK_k \Lp \frac{1}{\tau^2}\bA^{\sf T}\bA\bSigma, \frac{1}{\tau^2}\bA^{\sf T}\Lp \by - \bX\bbeta \Rp \Rp.$$
 The genHyBR algorithm creates an $n \times k$ basis $\bV_k = [\bv_1, \bv_2, \ldots, \bv_k]$ for this subspace, i.e., $\calS_k = \text{span } \{ \bV_1, \ldots, \bV_k \} $ using an efficient Golub-Kahan bidiagonalization iteration scheme which has been sketched in Algorithm \ref{tab:algo1}.

\begin{algorithm}
\begin{algorithmic}[1]
\ENSURE Matrices $\bA$, $\bSigma$, vector $\bb = \by - \bX\bbeta$ and $\tau^2$.
\STATE Compute $\bu_1 = \bb/\beta_1$, where $\beta_1 = \|\bb \|_{2}/\tau$.
\STATE Compute $\bv_1 = \frac{1}{\tau^2}\bA^{\sf T}\bu_1/\alpha_1$ where $\alpha_1 = \|\frac{1}{\tau^2}\bA^{\sf T}\bu_1 \|_{\bSigma}$.
\FOR {$i=1,\dots,k$}
\STATE Compute $\bu_{i+1} = \Lp \bA\bSigma\bv_i - \alpha_i\bu_i \Rp/\beta_{i+1}$ where $\beta_{i+1} = \frac{1}{\tau^2}\| \bA\bSigma\bv_i - \alpha_i\bu_i \|_{2}$.
\STATE Compute $\bv_{i+1} = \Lp \bA^{\sf T}\bu_{i+1}/\tau^2 - \beta_{i+1}\bv_i\Rp/\alpha_{i+1}$ where $\alpha_{i+1} = \| \bA^{\sf T}\bu_{i+1}/\tau^2 - \beta_{i+1}\bv_i \|_{\bSigma}$.
\ENDFOR
\RETURN $\beta_1, \bU_{k+1}, \bV_{k+1}$ and $\bB_k$.
\end{algorithmic}
\caption{Generalized Golub-Kahan (genGK) bidiagonalization}
\label{tab:algo1}
\end{algorithm}

From Algorithm \ref{tab:algo1}, we also obtain a $(k+1) \times k$ bidiagonal matrix $$\bB_k = \begin{bmatrix} \alpha_1 & & &\\
\beta_2 & \alpha_2 & &\\
& \ddots & \ddots&\\
& & \beta_k & \alpha_k\\
\end{bmatrix}.$$ 
The outputs of the algorithms satisfy the following relationships \beq \label{eq: exact_arith}\begin{split}
    &\bA \bSigma \bV_k = \bU_{k+1}\bB_k,\\
    &\bU_{k+1}^{\sf T}\bU_{k+1} = \tau^2 \bI_{k+1},\\
    &\bV_k^{\sf T}\bSigma\bV_k = \bI_k .
\end{split}  \eeq 

Since we are looking for a solution of $\bw \in \mathcal{S}_k$, we can write $\bw_k = \bV_k\bz_k$ and determine $\bz_k$ by solving \beq \begin{split}
    & \underset{\bw_k \in \mathcal{S}_k}{\text{ min }} \frac{1}{\tau^2}\|\bA\bSigma\bw_k - \bb\|_2^2 + \frac{1}{\sigma^2}\|\bw_k\|^2_{\bSigma}\\
    & \Leftrightarrow \underset{\bz_k \in \mathbb{R}^k}{\text{ min }} \|\bB_k\bz_k - \beta_1\be_1\|_2^2 + \frac{1}{\sigma^2}\|\bz_k\|^2_2 .
\end{split}  \eeq 
Therefore, given $\bB_k$ and $\bV_k$ and by undoing the change of variables, we approximate the solution to Eq. (\ref{eq: genHyBR_eq}) as \beq \label{eq: xhat_est} \bx_k^*(\btheta) = \bSigma \bV_k \Lp \bB_k^{\sf T}\bB_k + \frac{1}{\sigma^2}\bI \Rp^{-1}\bB_k^{\sf T}\beta_1 \be_1, \eeq where $\be_1$ is the first column of the $(k+1)\times (k+1)$ identity matrix; that is, the vector with the first entry $1$ and every other entry equal to $0$. In general, a stopping criterion must be used to terminate the iterations and to automatically determine the number of iterations $k$. Details on one such choice of stopping criterion are given in~\cite{chung2018efficient}. However, we do not use the said criterion for our method and instead treat the parameter $k$ as an algorithm parameter to be input by the user. The orthogonal basis vectors $\bu_k$ and $\bv_k$ may not remain numerically orthogonal and therefore may require a reorthogonalization scheme. Such a scheme is described in the \cite{chung2018efficient} paper and is available for the user to use in Kryging as well. However, we do not use it for the results presented in this paper.

The genHyBR method reduces the computational complexity of solving for $\bx$ from $\mathcal{O}(n^3)$ to that of matrix vector multiplication, $\mathcal{O}(n^2 + nk^2)$. When the latent process locations $\bs_1^*, \ldots, \bs_n^*$ are arranged on a rectangular grid, $\bSigma$ is symmetric BTTB and thus the matrix-vector multiplication can be achieved swiftly, in $\mathcal{O}(n \log n + nk^2)$ flops, using circulant embedding. Additionally, due to the form of $\bx_k^*(\btheta)$ in Eq. (\ref{eq: xhat_est}) and the exact arithmetic relationships presented in Eq. (\ref{eq: exact_arith}), the quadratic term $\widehat{\bx}(\btheta)^{\sf T}\bSigma^{-1}\widehat{\bx}(\btheta)$ can now be approximated as $\lVert \bz^*_k \rVert_2^2$, where $\bz^*_k = \Lp \bB_k^{\sf T}\bB_k + \frac{1}{\sigma^2}\bI \Rp^{-1}\bB_k^{\sf T}\beta_1 \be_1$. This requires only $\mathcal{O}(k^3)$ operations.

\subsection{Log-determinant Approximation}
\label{subsec:logdet}

To compute the $\log$ determinant of $\bSigma(\btheta)$, we once again use the symmetric BTTB structure of $\bSigma(\btheta)$.  \cite{gray2006toeplitz} reviews methods for creating a circulant matrix based on a Toeplitz matrix and using the circulant matrix structure to approximate the determinant of a Toeplitz matrix using inverse FFTs. Refer to Section $4.1, 4.4$ and $5.3$ of \cite{gray2006toeplitz} for details. This behavior can be extended to a BTTB structure as well and a similar asymptotic result also holds for them \citep{gyires1956eigenwerte, widom1974asymptotic}. The block circulant matrix $C = \Lp\Lp C_{jk} \Rp\Rp_{(2n_1-1) \times (2n_2-1)}$ can be created exactly as it is done for circulant embedding based matrix-vector product and therefore does not add any extra computation. The approximation to the log-determinant is of the form
\begin{equation*}
\begin{split}
    \logdethat \bSigma(\btheta) &= \sum_{p=1}^{n_1}\sum_{q=1}^{n_2} \log \Lp \sum_{j=1}^{2n_1-1}\sum_{k=1}^{2n_2-1} \right.\\
    & \left. \omega_{n_1}^{(j-1)(p-1)}  \omega_{n_2}^{(k-1)(q-1)} C_{jk} \Rp,
\end{split}
\end{equation*} where $\omega_{n_1} = \exp \Lp- 2\pi i/(2n_1-1)\Rp \text{ and } \omega_{n_2} = \exp \Lp- 2\pi i/(2n_2-1)\Rp$.


The approximation stems from the fact that the result is only exact in an asymptotic sense. However, numerical evidence suggests that the approximation to the $\log$ determinant and its derivatives improves as the number of grid points $n$ increases; a more precise statement of convergence can be found in Theorem 1.1 and Lemma 4.1(b) of~\cite{kent1996spectral}. We mention that besides the BCCB approximation, there are other ways of estimating the log-determinant, such as stochastic trace estimation~\citep{anitescu2012matrix,ubaru2017fast} and using Hierarchical matrix structure~\citep{ambikasaran2015fast,minden2017fast}.  In particular, the advantage of the stochastic trace estimator is that the information used in estimating the log-determinant can be reutilized during the computation of the gradient information. These approaches can be used in place of the proposed estimator.

\subsection{Optimization Details}
\label{subsec: optim_det}

The approximations described in the previous sections render the approximate profile log-likelihood function $\widetilde{\text{pl}}(\btheta)$ to have the form \beq \label{eq: prof_lik} \begin{split}
    \widetilde{\text{pl}}(\btheta) & \simeq -\frac{p}{2}\log \tau^2 -\frac{1}{2\tau^2}\bpsi^*_k(\btheta)^{\sf T}\bpsi^*_k(\btheta)\\
    & \quad- \frac{n}{2}\log \sigma^2 - \frac{1}{2}\logdethat (\bSigma(\btheta))\\
    & \quad - \frac{1}{2\sigma^2}\lVert \bz_k \rVert_2^2,
\end{split}  \eeq where $\bpsi^*_k = \by - \bX\bbeta - \bA\bx_k^*(\btheta)$,  $\bx_k^*(\btheta)$ and $\bz^*_k$ are as described in Section \ref{subsec:genHyBR} and $\logdethat (\bSigma(\btheta))$ is as described in Section \ref{subsec:logdet}. Evaluating this function is faster and we can put it in an optimization routine to optimize over $\btheta$ to get the estimates of $\btheta$ and $\widehat{\bx}(\btheta)$.

We use the MATLAB optimization routine \texttt{fminunc} with log-transformed range and variance parameters to avoid the non-negativity constrains. The optimization algorithm we use is a trust-region algorithm, which requires derivative information such as gradients and Hessians. The true gradient functions involve terms with $\bSigma(\btheta)^{-1}$ and therefore needs to be approximated. These problems are averted by using the genHyBR solution of $\bx_k^*(\btheta)$ in place of $\bx$ as the matrix inversion problem reduces to a matrix vector multiplication problem. The derivative of the $\log$ determinant is also approximated by using the BTTB structure. The details are given in Appendix~\ref{sec:gradhess}. To approximate the Hessian, we use a rank-one estimate of Hessian computed as the outer product of the approximate gradient. The rationale behind this approximation is the fact that, in expectation, the outer product of the score function equals the information matrix. Once again, the details are given in Appendix~\ref{sec:gradhess}. 

\subsection{Uncertainty Quantification}
\label{subsec:uq}

Besides a point estimate for $\bx$, we also want to quantify the uncertainty associated with the estimated $\bx$ and the predicted $\by$.  
We employ a parametric bootstrap for uncertainty quantification. Using the estimated $\hat{\btheta}$, we generate $B$ samples of $\bx_1, \ldots, \bx_B$ from a zero-mean Gaussian process. For each $\bx_b$, we generate $\by_b$ from the model in Eq. (\ref{eq: obs_mod}) with $\tau^2$ and $\bbeta$ replaced by their estimates. We then estimate $\hat{\by}_b$ by Kryging, but assuming $\btheta$ is known. 

On the set of prediction locations $\bs^*_1, \ldots, \bs_m^*$, we compute the bootstrap MSE for each location $\bs^*_i$ as \beq \label{eq: bootstrap_uq}
\begin{split}
\text{var }(x(\bs^*_i)|\hat{\btheta}) &\approx \frac{1}{B}\sum_{b=1}^B \Lp x_b(\bs^*_i|\hat{\btheta}) - \hat{x}_b(\bs^*_i|\hat{\btheta})\Rp^2,\\
\text{var }(y(\bs^*_i)|\hat{\btheta}) &\approx \frac{1}{B}\sum_{b=1}^B \Lp y_b(\bs^*_i|\hat{\btheta}) - \hat{y}_b(\bs^*_i|\hat{\btheta})\Rp^2.\\
\end{split}  \eeq This serves as an estimate of the classical Kriging variance for spatial prediction \citep{den2006correct}. Since we use a parametric bootstrap approach, we use $B=20$ bootstrap samples as just this many bootstrap samples provide satisfactory performance. The entire scenario entails using genHyBR method \citep{chung2018efficient} $B$ times, therefore costs $\mathcal{O}(n \log n + nk^2)$ flops. This procedure only approximates the uncertainty of the predictions assuming $\btheta$ is known.  However, the bootstrap could be extended to give standard errors for the elements of $\hat{\btheta}$ as well as prediction variances that account for uncertainty in $\btheta$ by simply estimating $\btheta$ for each bootstrap sample.

\subsection{Summary of the Method}
\label{subsec: sum_meth}

We now summarize the overall computational cost of this procedure. There are three main steps:
\begin{enumerate}
    \item Optimizing the profiled likelihood $\widetilde{\text{pl}}(\btheta)$ to obtain $\btheta^*$
    \item Compute $\bx_k^*(\btheta^*)$ and $\widehat{\by} = \bA\bx_k^*(\btheta^*)$.
    \item Compute prediction variance using bootstrap sampling.
\end{enumerate}
The optimization routine involves computing an approximate profile likelihood function and uses approximations based on the genGK algorithm to gradients and Hessian.  Using genGK algorithm takes only $\mathcal{O}(nk \log n)$ steps for computing $\bx_k^*(t)$ at the $t$-th iteration of the optimization.  

\textbf{Caveat:} Kryging depends on circulant embedding operations via the log-determinant approximation and bootstrap based uncertainty quantification. A successful execution requires that a positive definite embedding be found for the corresponding Gaussian process. Without this, the method may fail to produce a bootstrap sample from the Gaussian process in question and as a result fail to estimate uncertainty. This will also result in poor approximation of the log-determinant as many near-zero positive eigenvalues would be computed as near-zero negative eigenvalues and throw off the overall computation. This problem is evidently present when the spatial range parameter is high for the Gaussian process \citep[See ][]{graham2018analysis}. This problem with circulant embedding is well known. The problem of generating samples from a Gaussian process can be ameliorated by using different periodic embedding schemes \citep[See ][]{stein2002fast, gneiting2006fast, guinness2017circulant}. Forcefully resetting the small negative eigenvalues to zero or machine-precision value is a quick recourse for approximating the log-determinant. The different embedding schemes proposed in the literature may also be considered for this. However, none of these can solve the computational issue completely.

\section{Simulation Studies}
\label{sec:sim}


\setcounter{table}{0}

In this section, we perform simulation studies to evaluate the performance of our proposed method. These studies aim to demonstrate the effectiveness of the model with varying sample size as well as under different parametric settings for both gridded and irregularly spaced data. We perform three different simulation studies towards this goal. In each of the experiments, for each case, we repeat the process on 25 replications. Throughout the studies, the observed values $\by$ are created by adding noise to $\bx$, where $\bx$ is an observation from a Gaussian process with constant mean $\beta$ and exponential covariance function (i.e., $\Matern$ covariance with $\nu = 0.5$) with sill $\sigma^2$ and spatial range $\rho$. We take the variance of the noise process to be $\tau^2$.

The first study varies the number of observations $n$ by generating data on a $100 \times 100$, $200 \times 200$, $300 \times 300$ and $400 \times 400$ grid in the unit square. The covariate matrix $\bX$ is a single column vector of ones and the choice of $\btheta = (\beta,\sigma^2,\tau^2,\rho)^{\sf T}$ is taken to be $(44.49,3,0.5,0.1)$. The Kryging method is fit using the same grid of $p=n$ used to generate the data and we compare performance for $k\in\{20,50,100,200\}$. About $5\%$ of the observed data $\by$ were held out and were treated as test data upon which the performance was evaluated.

The second study demonstrates the performance of the method under different parametric settings on a grid of $200 \times 200$ points. The spatial extents were kept same as in the first study. The four different parametric settings that were used for this study are as follows: \begin{enumerate}
    \item Small spatial range, $\btheta = (44.49,3,0.5,0.05)^{\sf T}$.
    \item Large spatial range, $\btheta = (44.49,3,0.5,0.2)^{\sf T}$.
    \item Small partial sill, $\btheta = (44.49,1.5,0.5,0.1)^{\sf T}$.
    \item Large partial sill, $\btheta = (44.49,6,0.5,0.1)^{\sf T}$.
\end{enumerate} In all of these cases, about $5\%$ of the data from randomly chosen locations on the grid, were held out from the observed $\by$ and kept as test sample data on which to evaluate the method. 

The third study deals with the issue of irregularly spaced data. We used the first parametric setting, $\btheta = (44.49, 3, 0.5, 0.1)$ and the spatial extent of the data as in the first study.

The number of observed points were $40,000$ of which $5\%$ were held out as test samples. The data were generated by drawing $\bx$ on a $1000 \times 1000$ grid and discarding $96\%$ of the data at random, leaving an irregularly spaced dataset of $40,000$ observations. For testing the scalability with the grid size $n$, we used $200 \times 200$, $300 \times 300$ and $400 \times 400$ grids for $\bs^*_i$.

 The root mean squared error (RMSE) in predicting $\by$, pointwise coverage (CVG) of $95\%$ prediction intervals for these predictions were averaged over replications and median of computation time (MedTime) for all the replications were noted. These were used as performance metrics for each of the cases. For a competing method, we use the SPDE method available in the R package \texttt{INLA}. The SPDE method emerged from the comparison of several methods in \cite{heaton2019case} as one of the leading methods in terms of both computational speed and predictive accuracy.

\begin{table}[ht]
\centering
\caption{Table a) represents RMSE\textsubscript{Coverage} for predicting $\by$ over different grid sizes and different choices of the tuning parameter $k$ and the SPDE method, averaged over replications. The last column presents the maximum standard error for the given grid size across methods. Table b) shows the median computation times in minutes over different grid sizes and different choices of the tuning parameter $k$ and SPDE method. The figures in the bracket indicate standard errors.}
\label{tab: sim_1_rmsecov}
\flushleft \textbf{a)}\\
\centering
\begin{adjustbox}{max width=\linewidth}
\begin{tabular}{c|ccccc|c}
  \hline
  Grid Size & SPDE & \multicolumn{4}{c}{Kryging} & SE\\
  \cline{3-6}
 & & k=20 & k=50 & k=100 & k=200 & \\ 
  \hline
$100 \times 100$ & 0.91\textsubscript{0.95} & 0.93\textsubscript{0.92} & 0.91\textsubscript{0.91} & 0.91\textsubscript{0.91} & 0.91\textsubscript{0.91} & 0.03\textsubscript{0.03} \\ 
  $200 \times 200$ & 0.83\textsubscript{0.95} & 0.86\textsubscript{0.92} & 0.84\textsubscript{0.91} & 0.83\textsubscript{0.91} & 0.83\textsubscript{0.91} &  0.01\textsubscript{0.02} \\ 
  $300 \times 300$ & 0.80\textsubscript{0.95} & 0.86\textsubscript{0.92} & 0.84\textsubscript{0.91} & 0.83\textsubscript{0.91} & 0.83\textsubscript{0.91} &  0.01\textsubscript{0.02} \\ 
  $400 \times 400$ & 0.78\textsubscript{0.95} & 0.84\textsubscript{0.91} & 0.80\textsubscript{0.89} & 0.79\textsubscript{0.88} & 0.78\textsubscript{0.88} &  0.01\textsubscript{0.02} \\ 
   \hline
\end{tabular}
\end{adjustbox}\\
\flushleft \textbf{b)}\\
\centering
\begin{adjustbox}{max width=\linewidth}
\begin{tabular}{c|ccccc}
  \hline
 Grid Size & SPDE & \multicolumn{4}{c}{Kryging}\\
  \cline{3-6}
 & & k=20 & k=50 & k=100 & k=200\\ 
  \hline
$100 \times 100$ & 5.42 (0.66) & 0.14 (0.00) & 1.84 (0.49) & 5.14 (0.03) & 11.00 (0.15) \\ 
  $200 \times 200$ & 44.64 (10.57) & 5.51 (0.02) & 1.66 (0.28) & 2.12 (0.09) & 48.24 (0.54) \\ 
  $300 \times 300$ & 170.01 (21.46) & 3.39 (0.01) & 4.11 (0.03) & 5.49 (0.20) & 7.81 (0.24) \\ 
  $400 \times 400$ & 662.95 (108.14) & 10.78 (0.03) & 12.02 (0.12) & 14.28 (0.17) & 18.09 (0.22) \\ 
   \hline
\end{tabular}
\end{adjustbox}
\end{table}

Table \ref{tab: sim_1_rmsecov} presents the RMSE and pointwise coverage values, averaged over replications, for the first simulation study and the median time for computation over the replicates for different choices of the tuning parameter $k$ and different grid sizes. In all cases, $k=50$ seems to be sufficient. The occasional inconsistencies in the computation times in Table \ref{tab: sim_1_rmsecov} are due to the differences in the number of iterations taken by the optimization procedure to converge. In terms of RMSE and coverage, both the methods perform similarly but Kryging is considerably faster and is more scalable. On the other hand, the coverage for the proposed method is slightly below the nominal level. \textcolor{black}{This may be due to ignoring uncertainty in $\btheta$ when computing the prediction variances using Eq. \ref{eq: bootstrap_uq}. A possible fix for this is mentioned at the end of Section \ref{subsec:uq}. However, the coverage is not so low as to require such a fix sacrificing its fast runtime advantage.}

\begin{table}[ht]
\centering
\caption{Table a) represents RMSE\textsubscript{Coverage} for predicting $\by$ under different parametric settings for the SPDE and the proposed method with different choices of the tuning parameter $k$, averaged over replications. The last column presents the maximum standard error for the given setting across methods. Table b) shows median computation times in minutes over different choices of the tuning parameter $k$ and SPDE method for different parametric settings. The figures in the bracket indicate standard errors.} 
\label{tab: sim_2_rmsecov}
\flushleft \textbf{a)}\\
\centering
\begin{adjustbox}{max width=\linewidth}
\begin{tabular}{c|ccccc|c}
  \hline
  Setting & SPDE & \multicolumn{4}{c}{Kryging} & SE\\
  \cline{3-6}
 & & k=20 & k=50 & k=100 & k=200 & \\ 
  \hline
Setting 1 & 0.91\textsubscript{0.95} & 0.98\textsubscript{0.89} & 0.92\textsubscript{0.88} & 0.91\textsubscript{0.88} & 0.91\textsubscript{0.88} &  0.01\textsubscript{0.01} \\ 
  Setting 2 & 0.78\textsubscript{0.95} & 0.80\textsubscript{0.91} & 0.80\textsubscript{0.89} & 0.79\textsubscript{0.89} & 0.79\textsubscript{0.89} &  0.01\textsubscript{0.03} \\ 
  Setting 3 & 0.80\textsubscript{0.95} & 0.80\textsubscript{0.86} & 0.79\textsubscript{0.83} & 0.79\textsubscript{0.83} & 0.79\textsubscript{0.83} &  0.08\textsubscript{0.03} \\ 
  Setting 4 & 0.90\textsubscript{0.95} & 0.98\textsubscript{0.96} & 0.92\textsubscript{0.96} & 0.91\textsubscript{0.96} & 0.90\textsubscript{0.96} &  0.02\textsubscript{0.01} \\ 
   \hline
\end{tabular}\end{adjustbox}\\
\flushleft \textbf{b)}\\
\centering
\begin{adjustbox}{max width=\linewidth}
\begin{tabular}{c|ccccc}
  \hline
 Setting & SPDE & \multicolumn{4}{c}{Kryging}\\
  \cline{3-6}
 & & k=20 & k=50 & k=100 & k=200\\ 
 \hline
Setting 1 & 17.69 (2.63) & 0.90 (0.00) & 1.40 (0.08) & 2.15 (0.10) & 48.25 (0.47) \\ 
  Setting 2 & 19.03 (1.06) & 5.56 (0.05) & 1.60 (0.25) & 2.12 (0.07) & 48.74 (0.88) \\ 
  Setting 3 & 18.26 (3.10) & 5.47 (0.41) & 1.37 (0.08) & 2.17 (0.08) & 48.53 (0.27) \\ 
  Setting 4 & 18.49 (2.19) & 5.57 (0.04) & 2.37 (1.02) & 2.16 (0.08) & 48.62 (0.52) \\ 
   \hline
\end{tabular}\end{adjustbox}
\end{table}

The results for the $200 \times 200$ grids with different true spatial covariance parameters are given in Table \ref{tab: sim_2_rmsecov}.  For Settings 2 and 3, $k=25$ works well.  This is not surprising for Setting 2 because  the process with large range is smooth as easier to represent with a small number of terms.  Solid performance for small $k$ in Setting 3 with lower partial sill is also expected because genHyBR \citep{chung2018efficient} makes use of the partial sill to nugget ratio being moderate. As in the first simulation, going beyond $k=50$ seems unnecessary and the prediction RMSE performance is comparable to that of the SPDE method, but with substantially faster computation. \textcolor{black}{Since \texttt{INLA} is implemented in R and Kryging is implemented in Matlab, the difference in platform makes the computing time comparisons difficult to interpret. However, the gain in computation time for Kryging is likely not the result of change in platform solely because \texttt{INLA} is highly optimized code \citep{martino2009implementing}.}

\begin{table}[ht]
\centering
\caption{Table a) represents RMSE\textsubscript{Coverage} for predicting $\by$ for the SPDE and Kryging with different choices of the tuning parameter $k$ and different underlying gridsizes, averaged over replications for irregularly spaced datasets. The last column presents the maximum standard error for the given setting across methods. Table b) shows median computation times in minutes over different grid sizes and different choices of the tuning parameter $k$ and SPDE method. The figures in the bracket indicate standard errors.}
\label{tab: sim_irreg}
\flushleft \textbf{a)}\\
\centering
\begin{adjustbox}{max width=\linewidth}
\begin{tabular}{c|cccc|c}
\hline
  Gridsize & SPDE & \multicolumn{3}{c}{Kryging} & SE\\
  \cline{3-5}
 &  & k=20 & k=50 & k=100 &  \\ 
  \hline
$200 \times 200$ & 0.82\textsubscript{0.95} & 0.85\textsubscript{0.90} & 0.83\textsubscript{0.88} & 0.83\textsubscript{0.87} & 0.02\textsubscript{0.02} \\ 
 $300 \times 300$ & 0.82\textsubscript{0.95} & 0.85\textsubscript{0.90} & 0.83\textsubscript{0.89} & 0.83\textsubscript{0.88} & 0.02\textsubscript{0.02} \\ 
 $400 \times 400$ & 0.82\textsubscript{0.95} & 0.85\textsubscript{0.91} & 0.83\textsubscript{0.89} & 0.82\textsubscript{0.89} & 0.02\textsubscript{0.02} \\ 
   \hline
 
\end{tabular}\end{adjustbox}
\flushleft \textbf{b)}\\
\centering
\begin{adjustbox}{max width=\linewidth}
\begin{tabular}{c|cccc}
\hline
 Gridsize & SPDE & \multicolumn{3}{c}{Kryging}\\
  \cline{3-5}
 &  & k=20 & k=50 & k=100\\ 
  \hline
 $200 \times 200$ & 33.55 (3.58) & 8.08 (0.02) & 3.33 (0.31) & 3.93 (0.15) \\ 
 $300 \times 300$ & 33.55 (3.58) & 4.86 (0.02) & 5.58 (0.03) & 6.85 (0.05) \\ 
 $400 \times 400$ & 33.55 (3.58) & 8.58 (0.10) & 9.96 (0.11) & 12.40 (0.09) \\ 
  \hline
\end{tabular}\end{adjustbox}
\end{table}

The results for irregularly-spaced data are shown in Table \ref{tab: sim_irreg}. The performance is similar to the SPDE method for the proposed method with slight undercoverage. In essence, the performance is quite similar to the regularly gridded data scenario in the first simulation study. 

We also check the performance of the proposed method in estimating the true mean and spatial covariance parameters against those obtained from SPDE. Across all settings and irrespective of whether the data was on a regular grid or not, the results are consistent. While SPDE does a better job at estimating the nugget parameter, Kryging does a better job at estimating the partial sill. For estimating range and the mean parameters, both the method perform similarly. Detailed comparisons are presented in tables in Appendix \ref{app:addl_tab}.

\section{Application to MODIS/Terra Land Surface Temperature Data}
\label{sec:apply}


In this section, we analyze a real dataset using the proposed method. We use the dataset used by \cite{heaton2019case} for a comparison of methods for analyzing massive spatial data. The dataset consists of Level-3 data on land surface temperatures as measured by the Terra instrument onboard the MODIS satellite on August 4, 2016. The original data was available in MODIS reprojection tool web (MRTweb) which has since been decomissioned. The entire dataset is available in the GitHub repository for the \cite{heaton2019case} project at \href{https://github.com/finnlindgren/heatoncomparison}{this GitHub repository}. The main reason for using this dataset is so that we can compare to other existing methods easily as this dataset was previously analyzed by twelve other existing methods in \cite{heaton2019case}.

\begin{figure}
    \includegraphics[width=0.9\linewidth]{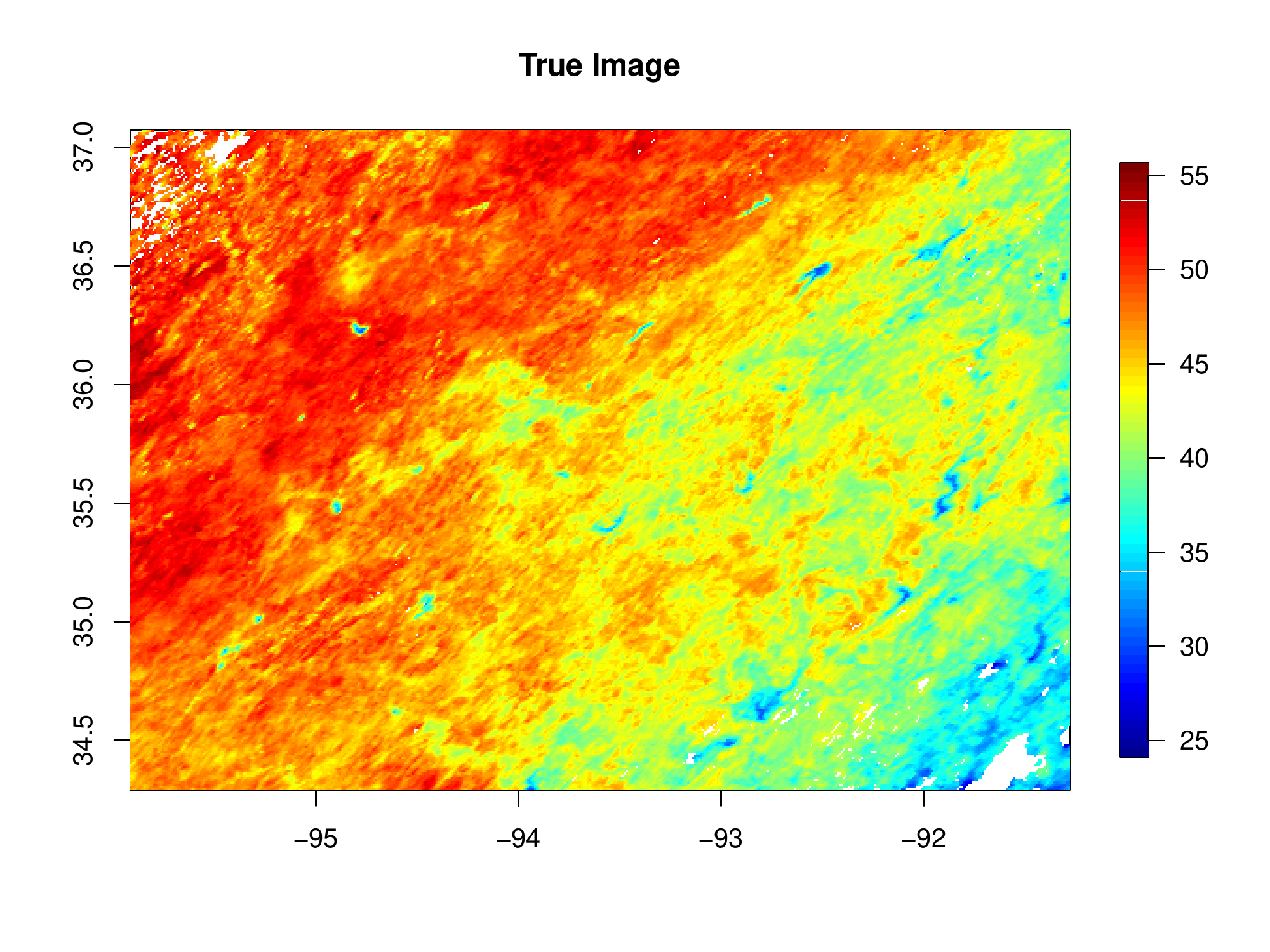}    \includegraphics[width=0.9\linewidth]{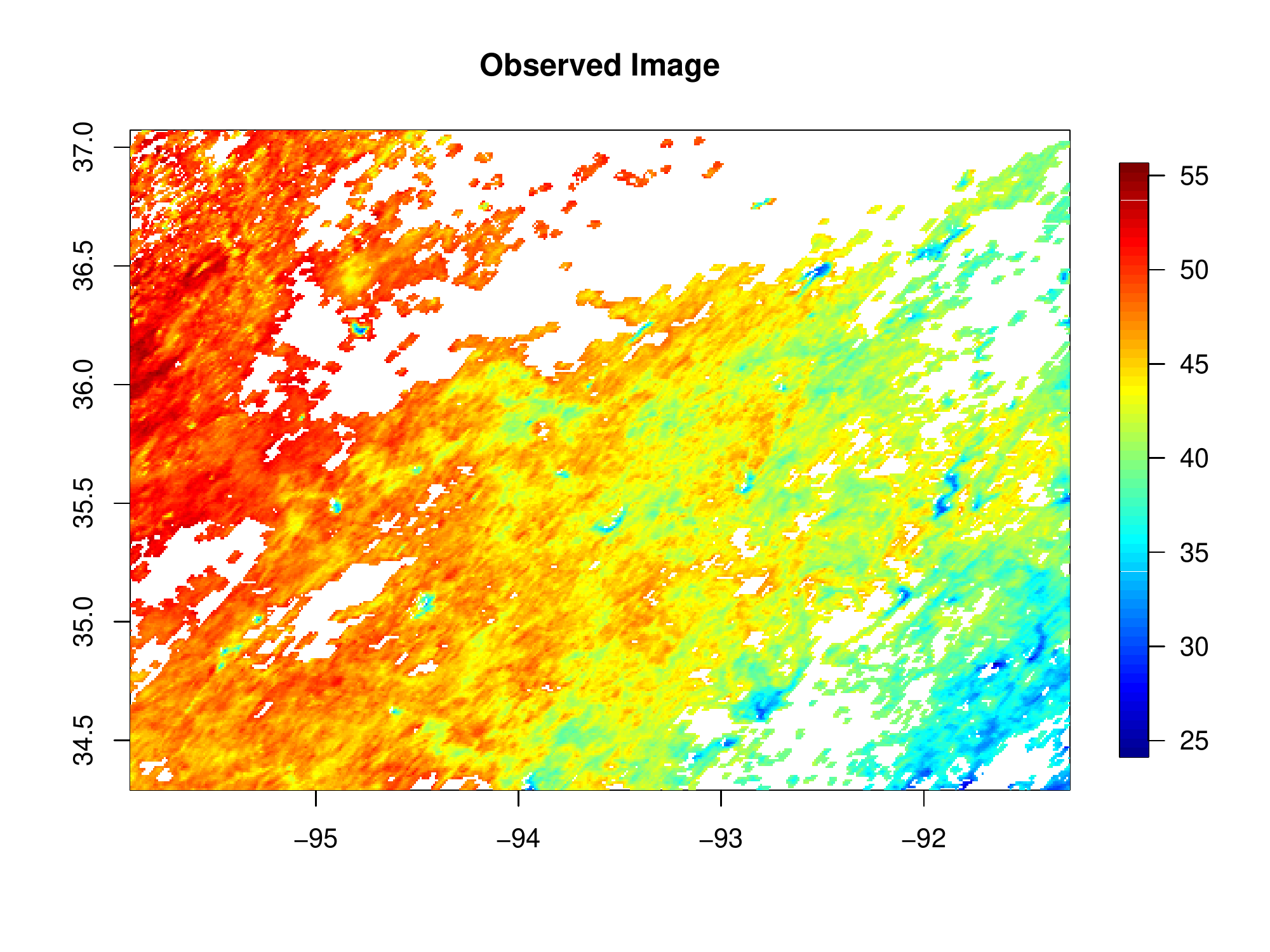}
    \includegraphics[width=0.9\linewidth]{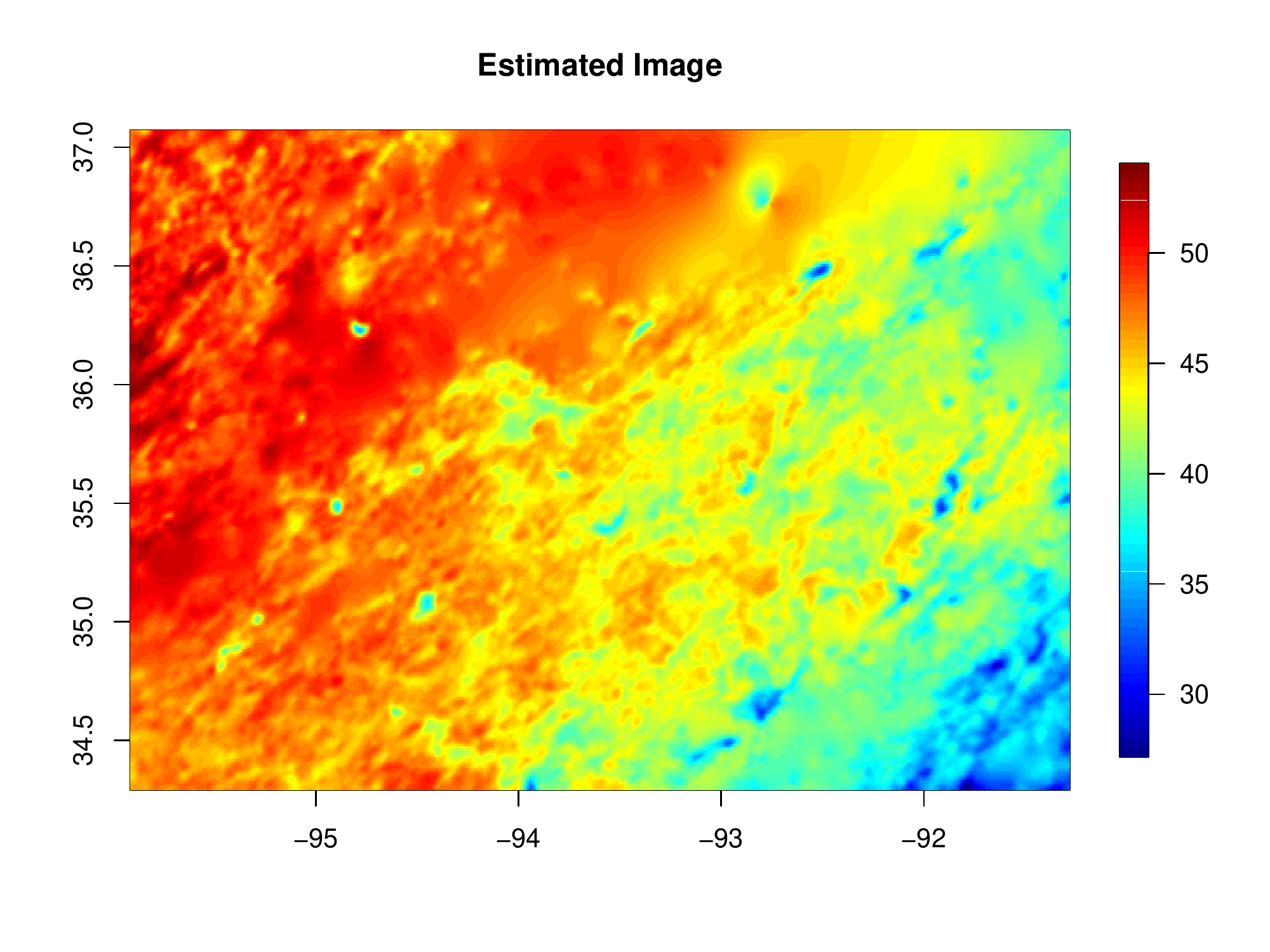}    \includegraphics[width=0.9\linewidth]{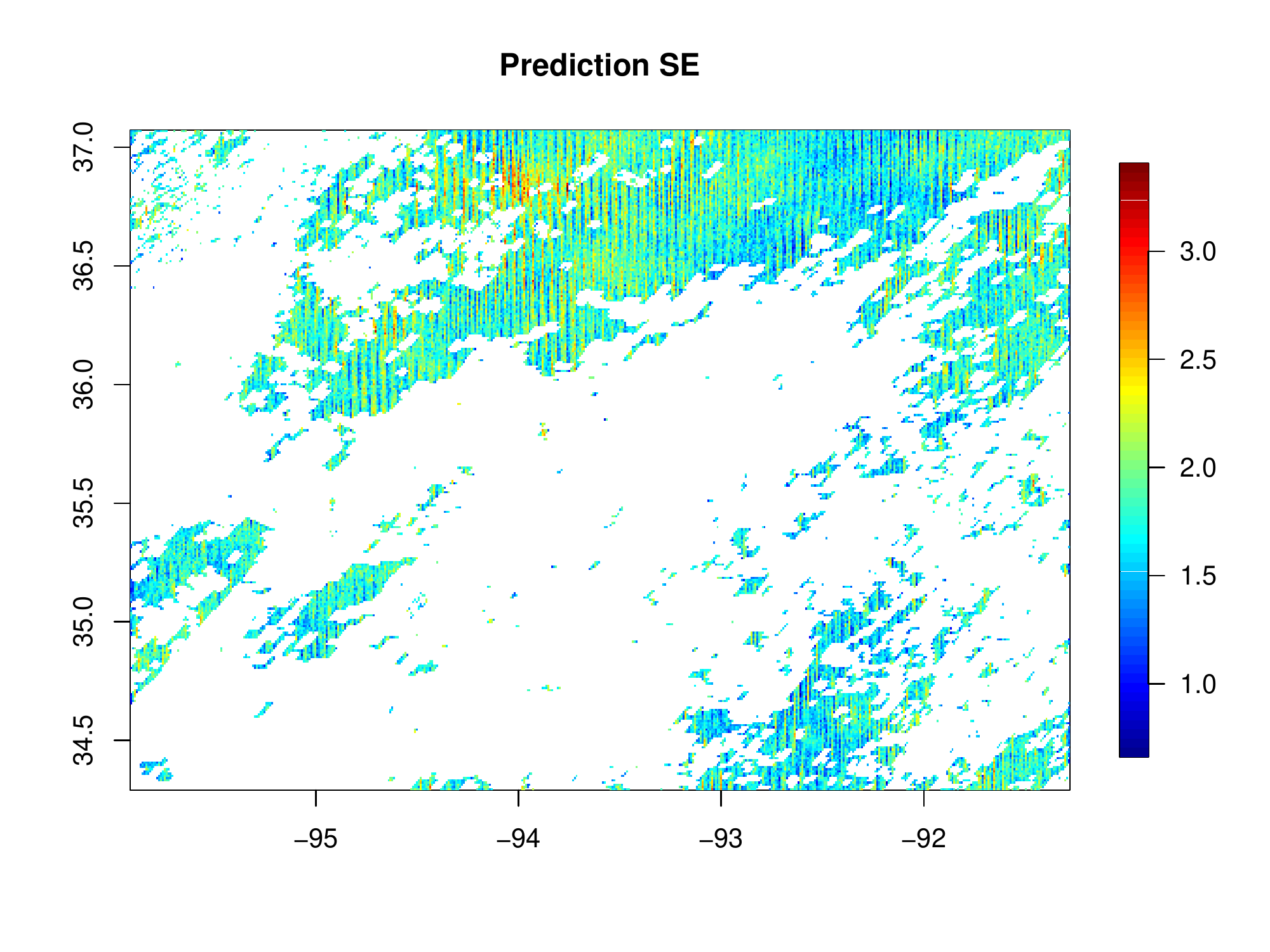}
    \caption{True satellite image (top), the image used for training after holding out data for test sample (second top), the image obtained from the estimated values (second bottom) and the prediction standard errors (bottom) for $k=200$.}
    \label{fig:data_app}
\end{figure}

The observations were laid out on a regular grid of size $500 \times 300$ within longitude values $-95.91153$ to $-91.28381$ and latitude values $34.29519$ to $37.06811$. About $1.1\%$ of the data, $1,691$ grid cells out of $150,000$ cells, were corrupted due to cloud cover. A further $42,740$ observations were held out from the training set, keeping about $70\%$ of the data in the training set and about $30\%$ in the test set. The training and testing datasets along the locations are available in the previously mentioned \href{https://github.com/finnlindgren/heatoncomparison}{GitHub repository}. Figure \ref{fig:data_app} shows the true data (top) and training data (second top) created after removing some observations.

We ran the Kryging algorithm with  $k = 50$, $100$, $200$ and $300$. For each value of $k$, we use different initial values and pick the best one using five-fold cross-validation within the training dataset. The mean absolute error (MAE), root mean squared error (RMSE), continuously ranked probability score (CRPS), interval score (INT) and pointwise coverage (CVG) for the predictions of the test set datapoints were computed for each case and the computation times were noted and are tabulated in Table \ref{tab:data_res}. Figure \ref{fig:data_app} shows the estimates (bottom left) and corresponding standard errors (bottom right) for the data. The estimated image picks up all the spatial features in the true data, indicating a good fit. Since the same dataset was also analyzed by twelve other existing methods, the above mentioned metrics for which are available in the \cite{heaton2019case} paper. The relevant results, as presented in the original paper, are presented in Table \ref{tab:heaton_res}. This allows us a chance to compare the performance of our method to other existing methods, although the computing platforms were not the same for the two cases.

\begin{table}
    \centering
    \caption{Performance of the proposed method on the MODIS dataset for various choices of $k$.}
    \label{tab:data_res}
    \begin{adjustbox}{max width=\linewidth}
    \begin{tabular}{cccccccc}
    \hline
         $k$ & MAE & RMSE & CRPS & INT & CVG & Run time (min.) & Cores used \\
         \hline
         $50$&1.43&1.95&1.07&10.97&0.93&11.18&4\\
         $100$&1.43&1.85&1.04&9.74&0.93&15.10&4\\
         $200$&1.36&1.78&0.99&9.60&0.93&19.59&4\\
         $300$&1.36&1.79&0.99&9.68&0.93&27.01&4\\
         \hline
    \end{tabular}\end{adjustbox}
\end{table}

\begin{table}
    \centering
    \caption{Results from the case study competition for the satellite data as in Table 3 of \cite{heaton2019case}.}
    \label{tab:heaton_res}
    \begin{adjustbox}{max width=\linewidth}
    \begin{tabular}{c|ccccccc}
         \hline
         Method & MAE & RMSE & CRPS & INT & CVG & Run time(min) & Cores used\\
         \hline
         FRK&1.96&2.44&1.44&14.08&0.70&2.32&1\\
         Gapfill&1.33&1.86&1.17&34.78&0.36&1.39&40\\
         LatticeKrig&1.22&1.68&0.87&7.55&0.96&27.92&1\\
         LAGP&1.65&2.08&1.17&10.81&0.83&2.27&40\\
         Metakriging&2.08&2.50&1.44&10.77&0.89&2888.52&30\\
         MRA&1.33&1.85&0.94&8.00&0.92&15.61&1\\
         NNGP&1.21&1.64&0.85&7.57&0.95&2.06&10\\
         Partition&1.41&1.80&1.02&10.49&0.86&79.98&55\\
         Pred. Proc.&2.15&2.64&1.55&15.51&0.83&160.24&10\\
         SPDE&1.10&1.53&0.83&8.85&0.97&120.33&2\\
         Tapering&1.87&2.45&1.32&10.31&0.93&133.26&1\\
         Periodic Embedding&1.29&1.79&0.91&7.44&0.93&9.81&1\\
         \hline
    \end{tabular}\end{adjustbox}
\end{table}

 In terms of RMSE and coverage, SPDE \citep{lindgren2011explicit}, Nearest Neighbor Gaussian Process  or NNGP \citep{datta2016hierarchical,datta2016nearest,datta2016nonseparable} and LatticeKrig \citep{nychka2015multiresolution} perform better than the proposed method. The time taken by the method is significantly less than the SPDE method and comparable to LatticeKrig. \textcolor{black}{Although it should be mentioned that the they were run in different platforms with similar hardware set-up, so the comparison should not be considered a direct one.} The time presented for the NNGP method in \cite{heaton2019case} considers only the time taken for the conjugate model where a well defined grid of possible parameter values were supplied to the model to use cross-validation in parallel. This range of parameter values need to be determined first and is the more difficult and time consuming part of any existing approximate inference method and neither the strategy nor the time taken to arrive at those numbers were reported in \cite{heaton2019case}.

\section{Conclusion}
\label{sec:conc}


In this article, we propose an approximate inference method for analyzing massive spatial datasets using Krylov subspace approximation and profile maximum likelihood methods. The method assumes that the underlying process realizations are on a regular equispaced grid, but the observations need not be colocated on the grid.  While we exclusively model the spatial process covariance using the  $\Matern$ covariance family, the method works for any choice of stationary covariance function. We also propose an approach to approximate log-determinants for symmetric BTTB matrices which has guaranteed asymptotic convergence to the true log-determinant value. The method has computational complexity of $\mathcal{O}(n \log n)$, resulting in fast run times and excellent scalability with the sample size while producing decent estimates and requires little tuning. The method is expected to run especially well when the spatial range is small to moderate and partial sill to nugget ratio is moderate. This is seen in the applications involving both synthetic and real datasets. 

Although uncertainties for the mean and spatial parameter estimates are not provided directly, they can be obtained using the following approaches. A reasonable approach would be to compute the exact Hessian and its inverse for the optimization process of Eq. (\ref{eq: prof_lik}). However, that is time consuming as it has $\mathcal{O}(n^3)$ complexity involved with the computation. A suitable approximation to the inverse of the Hessian will be needed to efficiently estimate the uncertainties associated with these parameters. A computationally-expensive alternative is to estimate the parameters using the parametric bootstrap, as outlined in Section \ref{subsec:uq}.

\textcolor{black}{The method is proposed as a $d$-dimensional method. However, for irregular datasets on dimensions higher than 3, the grid formation is slow and difficult. But for the purposes of geostatistical analyses, we need only concern ourselves with problems in $\rtwo$ or $\rtwo \times \rone$ where grids are simple and easy to deal with. Should the case arise where one has to deal with higher dimensional geospatial analysis, one needs to look for a suitable alternative to the grid structure which can be a future avenue for research. Moreover, Kryging is most attractive when the observations are approximately on a grid or uniformly distributed and adaptations for extremely irregular cases such as data observed along transects or in separated clusters is another area of future work.}

The proposed model can be utilized in many other scenarios than simply what has been illustrated in this article. The computational amenities of the method can be utilized for spatiotemporal modeling. Changing the observational model to include two or more sources of data can be contemplated as well. Quantifying uncertainties for the mean and the spatial parameters can be one possible extension. Extending the method to non-Gaussian observational models, for example, binary or count data, would be another possibility.

\section*{Acknowledgements}

The authors were partially supported by the National Science Foundation through the awards DMS-1845406 and DMS-1638521. The authors were also partially supported by the National Institute of Health through the awards\\ R01ES031651-01 and R01ES027892 and by The King Abdullah University of Science and Technology grant 3800.2. We would like to thank them for their support.

\section*{Declarations}
\subsection*{Funding}
The authors were partially supported by the National Science Foundation through the awards DMS-1845406 and DMS-1638521. The authors were also partially supported by the National Institute of Health through the awards\\ R01ES031651-01 and R01ES027892 and by The King Abdullah University of Science and Technology grant 3800.2.
\subsection*{Conflict of Interest}
The authors have no conflicts of interest to declare that are relevant to the content of this article.

\subsection*{Availability of Data and Material}
The dataset analyzed in Section \ref{sec:apply} is available in the GitHub repository for the \cite{heaton2019case} project at \href{https://github.com/finnlindgren/heatoncomparison}{this GitHub repository}. 

\subsection*{Code Availability}
A \href{https://github.com/SumanM47/Krylov_Approx.git}{GitHub repository} has been set up that contains codes and a demonstration file for the methods described in the article.

\subsection*{Ethics Approval}
Not Applicable

\subsection*{Consent to Participate}
Not Applicable

\subsection*{Consent for Publication}
Not Applicable

\bibliographystyle{spbasic}
\bibliography{Bibliography-MM-MC}

\appendix

\begin{center}
    \textbf{APPENDIX}
\end{center}

\section{Gradient and Hessian Computation for the Optimization Procedure}\label{sec:gradhess}

In this section, we present the necessary details of computing and approximating the gradient and Hessian for the optimization routine. 

We first, derive exact expressions for the gradient and then show how to approximate them using the strategy in Sections~\ref{subsec:genHyBR} and~\ref{subsec:prof_lik}. Computing the analytical gradient would require computing derivatives of $\bGamma = \Lp \frac{1}{\sigma^2}\bSigma(\btheta)^{-1} + \frac{1}{\tau^2}\bA^{\sf T}\bA \Rp^{-1}$ and $\widehat{\bx}(\btheta)$ with respect to each of $\mu, \sigma^2, \tau^2$ and $\rho$. For convenience, we reparametrize $1/\sigma^2 = \lambda^2$ and $1/\tau^2 = \lambda_e^2$. Using the precision instead of variance brings about greater ease in computing the analytical derivatives. Under the new parametrization, \beq \label{eq: gamma_repar} \bGamma = \Lp \lambda_e^2\bA^{\sf T}\bA + \lambda^2\bSigma^{-1} \Rp^{-1}, \eeq \beq \label{eq: x_act_repar} \widehat{\bx}(\btheta) = \bGamma \lambda_e^2\bA^{\sf T}(\by-\bX\bbeta), \eeq and \beq \label{eq: exact_plik_repar}
\begin{split}
    \text{pl}(\btheta) & \simeq \frac{p}{2}\log \lambda_e^2 -\frac{\lambda_e^2}{2}\widehat{\bpsi}(\btheta)^{\sf T}\widehat{\bpsi}(\btheta)\\
    & \quad + \frac{n}{2}\log \lambda^2 - \frac{1}{2}\logdet\bSigma(\btheta) -\\ 
    & \quad \frac{\lambda^2}{2}\hat{\bx}(\btheta)^{\sf T}\bSigma(\btheta)^{-1}\hat{\bx}(\btheta),
\end{split} \eeq where $\widehat{\bpsi}(\btheta) = \by - \bX\bbeta - \bA\hat{\bx}(\btheta)$.

The derivatives for $\bGamma$ are computed to be \beq \label{eq: gamma_der}  \begin{split}
    \frac{\partial \bGamma}{\partial \bbeta} & = \mathbf{0},\\
    \frac{\partial \bGamma}{\partial \lambda^2} & = - \bGamma\bSigma(\rho)^{-1}\bGamma,\\
    \frac{\partial \bGamma}{\partial \rho} & = \lambda^2 \bGamma \bSigma(\rho)^{-1}\Lp \text{d}\bSigma(\rho) \Rp\bSigma(\rho)^{-1} \bGamma, \\
    \frac{\partial \bGamma}{\partial \lambda_e^2} & = - \bGamma \bA^{\sf T}\bA \bGamma,
\end{split} \eeq where $\text{d}\bSigma(\rho)$ denotes the derivative of $\bSigma(\rho)$ with respect to $\rho$. This is easy to compute analytically and has the nice BTTB property that $\bSigma(\rho)$ has.

Using the expressions in (\ref{eq: gamma_der}), we compute the derivatives of $\widehat{\bx}(\btheta)$ to be \beq \label{eq: x_der} \begin{split}
    \frac{\partial \widehat{\bx}(\btheta)}{\partial \bbeta} & = -\lambda_e^2 \bGamma \bA^{\sf T}\bX,\\
    \frac{\partial \widehat{\bx}(\btheta)}{\partial \lambda^2} & = -\bGamma \bSigma(\rho)^{-1} \widehat{\bx}(\btheta),\\
    \frac{\partial \widehat{\bx}(\btheta)}{\partial \rho} & = \lambda^2 \bGamma \bSigma(\rho)^{-1} \Lp \text{d} \bSigma(\rho)\Rp \bSigma(\rho)^{-1}\widehat{\bx}(\btheta),\\
    \frac{\partial \widehat{\bx}(\btheta)}{\partial \lambda_e^2} & = \bGamma\bA^{\sf T}\widehat{\bpsi}(\btheta) .
\end{split} \eeq

Substituting the expressions for analytical derivatives of $\bGamma$ and $\widehat{\bx}(\btheta)$ in the expression for the analytical gradient, we have it computed to be \beq \label{eq: anal_grad} \begin{split}
          \frac{\partial \text{pl}}{\partial \bbeta} & = \> \lambda_e^2\bX^{\sf T}\widehat{\bpsi}(\btheta),\\
        \frac{\partial \text{pl}}{\partial \lambda^2} & = \>\frac{n}{2\lambda^2} -  \frac{1}{2}\widehat{\bx}(\btheta)^{\sf T}\bSigma(\rho)^{-1}\widehat{\bx}(\btheta)\\
        \frac{\partial \text{pl}}{\partial \rho} & = -\frac{1}{2}\text{dL } + \frac{1}{2}\lambda^2 \widehat{\bx}(\btheta)^{\sf T}\bSigma(\rho)^{-1}\Lp \text{d}\bSigma(\rho) \Rp\bSigma(\rho)^{-1}\widehat{\bx}(\btheta),\\
        \frac{\partial \text{pl}}{\partial \lambda_e^2} & = \> \frac{p}{2\lambda_e^2} - \frac{1}{2}\widehat{\bpsi}(\btheta)^{\sf T}\widehat{\bpsi}(\btheta),
\end{split}\eeq where $\text{dL}$ is the derivative of $\logdet \bSigma(\rho)$ with respect to $\rho$.

We approximate the gradient expressions in (\ref{eq: anal_grad}) by approximating $\widehat{\bx}(\btheta)$ by $\bx_k^*(\btheta)$ as in (\ref{eq: xhat_est}) and using the exact arithmetic identities expressed in (\ref{eq: exact_arith}). The approximated gradients can be computed as \beq \label{eq: approx_grad} \begin{split}
    \frac{\partial \text{pl}}{\partial \bbeta} & \approx \lambda_e^2 \bX^{\sf T}\bpsi^*_k(\btheta),\\
    \frac{\partial \text{pl}}{\partial \lambda^2} & \approx \frac{n}{2\lambda^2} - \frac{1}{2}\|\mathbf{z}_k\|_2^2,\\
    \frac{\partial \text{pl}}{\partial \rho} & \approx -\frac{1}{2}\widehat{\text{dL}}  + \frac{\lambda^2}{2}\bz_k^{\sf T}\bV_k^{\sf T}\Lp \text{d}\bSigma(\rho) \Rp\bV_k\bz_k,\\
    \frac{\partial \text{pl}}{\partial \lambda_e^2} & \approx \frac{p}{2\lambda_e^2} - \frac{1}{2}\bpsi^*_k(\btheta)^{\sf T}\bpsi^*_k(\btheta),
\end{split} \eeq where $\bpsi^*_k(\btheta) = \by - \bX\bbeta - \bA\bx^*_k(\btheta)$ and $\bV_k, \bz_k$ have been defined in Section~\ref{subsec:genHyBR}.

$\widehat{\text{dL}}$ is an approximation to $\text{dL}$, the derivative of the log-determinant of $\bSigma(\rho)$ with respect to $\rho$. The analytical expression for $\text{dL}$ turns out to be \[\text{dL} = \text{trace}\Lp\bSigma(\rho)^{-1}\text{d}\bSigma(\rho).\Rp\] This is infeasible to compute directly and is therefore approximated using the BTTB structure of $\bSigma(\rho)$ and $\text{d}\bSigma(\rho)$. 

Any symmetric matrix with BTTB structure can be extended to have a BCCB structure as was done in computing the log-determinant itself and one can extract the eigenvalues of the matrix with BTTB structure using the matrix with BCCB structure. Any BCCB matrix is diagonalizable as $\bF\bD\bF^{\sf T}$, where $\bF$ is a scaled matrix consisting of $d$-dimensional (d=2, in our case) Fourier coefficients, irrespective of the BCCB matrix being diagonalized. Therefore, we can say \beq \label{eq: circ_diag}
\begin{split}
    \bSigma(\rho) &= \bF \bD_1 \bF^{\sf T},\\
    \bSigma(\rho)^{-1} &= \bF \bD_1^{-1} \bF^{\sf T},\\
    \text{d}\bSigma(\rho) & = \bF \bD_2 \bF^{\sf T}.
\end{split}
\eeq These imply that \beq \label{eq: trace_approx} \begin{split}
    \text{trace}\Lp \bSigma(\rho)^{-1}\text{d}\bSigma(\rho) \Rp & = \text{trace}\Lp \bF \bD_1^{-1}\bF^{\sf T}\bF \bD_2 \bF^{\sf T} \Rp\\
    & \quad = \text{trace}\Lp \bD_1^{-1}\bD_2 \Rp.
\end{split}
\eeq Since both $\bD_1$ and $\bD_2$ are diagonal, approximating $\text{dL}$ boils down to computing $\bD_1$ and $\bD_2$ which can be computed by $d$-dimensional FFT of the corresponding first circulant block structures of the extended BCCB structure and subsetting it properly. The equivalence in computing the derivative of log-determinant of the BTTB and matrix and its corresponding BCCB matrix has been demonstrated by \cite{kent1996spectral}, showing the approximation to have the same error rate as in approximating the log-determinant itself. Approximating the derivative of the log-determinant term also costs the same as approximating the log-determinant itself, $\mathcal{O}(n \log n)$.

While minimizing the negative log-likelihood function, the Hessian turns out to be simply the Information matrix $\bI(\btheta)$. While \[\mathbb{E}\Lp -\nabla_2 \text{pl}(\btheta) \Rp = \bI(\btheta),\] we also have \[\mathbb{E}\Lp\nabla \text{pl}(\btheta) \nabla \text{pl}(\btheta)^{\sf T} \Rp = \mathbb{E}\lp\Lp-\nabla \text{pl}(\btheta)\Rp \Lp -\nabla \text{pl}(\btheta)\Rp^{\sf T} \rp.\] Here the expectations are computed with respect to $\by$ and $\nabla$, $\nabla_2$ represent the gradient and Hessian created by computing first and second order partial derivatives with respect to $\btheta$. Therefore, the outer product of the gradient with itself serves as a rank-one estimate for the Hessian for a likelihood optimization problem. Although we are using profile likelihood instead of the actual likelihood function, the approximation still stands in an asymptotic sense since both the actual likelihood estimator and the profile likelihood estimators have the same asymptotic properties. This prompts us to take the outer product of the approximated gradient with itself as a rank-one approximation to the Hessian.

However, we compute the unique entries of the exact Hessian to be \beq \label{eq: Hess_ex}
\begin{split}
    \frac{\partial^2 \text{pl}}{\partial \bbeta \partial \bbeta^{\sf T}} &= -\lambda_e^2\bX^{\sf T}\bX + \lambda_e^4\bX^{\sf T}\bA\bGamma\bA^{\sf T}\bX\\
    \frac{\partial^2 \text{pl}}{\partial \bbeta \partial \lambda^2} &= \lambda_e^2\bX^{\sf T}\bA\bGamma\bSigma(\rho)^{-1}\widehat{\bx}(\btheta)\\
    \frac{\partial^2 \text{pl}}{\partial \bbeta \partial \rho} &= -\lambda_e^2\lambda^2\bX^{\sf T}\bA\bGamma\bSigma(\rho)^{-1}\Lp \text{d}\bSigma(\rho) \Rp \bSigma(\rho)^{-1}\widehat{\bx}(\btheta)\\
    \frac{\partial^2 \text{pl}}{\partial \bbeta \partial \lambda_e^2} &= \bX^{\sf T}\widehat{\bpsi}(\btheta) - \lambda_e^2\bX^{\sf T}\bA\bGamma\bA^{\sf T}\widehat{\bpsi}(\btheta)\\
    \frac{\partial^2 \text{pl}}{\partial \lambda^4} &= -\frac{n}{2\lambda^4} + \widehat{\bx}(\btheta)^{\sf T}\bSigma(\rho)^{-1}\bGamma\bSigma(\rho)^{-1}\widehat{\bx}(\btheta)\\
    \frac{\partial^2 \text{pl}}{\partial \lambda^2 \partial \rho} &= \frac{1}{2}\widehat{\bx}(\btheta)^{\sf T}\bSigma(\rho)^{-1}\Lp \text{d}\bSigma(\rho) \Rp\bSigma(\rho)^{-1}\widehat{\bx}(\btheta)\\
    & \quad - \lambda^2\widehat{\bx}(\btheta)^{\sf T}\bSigma(\rho)^{-1}\Lp \text{d}\bSigma(\rho) \Rp\bSigma(\rho)^{-1}\bGamma\bSigma(\rho)^{-1}\widehat{\bx}(\btheta)\\
    \frac{\partial^2 \text{pl}}{\partial \lambda^2 \partial \lambda_e^2} &= -\widehat{\bx}(\btheta)^{\sf T}\bSigma(\rho)^{-1}\bGamma\bA^{\sf T}\widehat{\bpsi}(\btheta)\\
    \frac{\partial^2 \text{pl}}{\partial \rho^2} &= -\frac{1}{2}\text{d}^2\text{L} + \frac{\lambda^2}{2}\widehat{\bx}(\btheta)\bSigma(\rho)^{-1}\Lp \text{d}^2\bSigma(\rho) \Rp \bSigma(\rho)^{-1}\widehat{\bx}(\btheta)\\
    & \quad - \lambda^2\widehat{\bx}(\btheta)^{\sf T}\bSigma(\rho)^{-1}\Lp \text{d}\bSigma(\rho) \Rp \lp \bSigma(\rho)^{-1}\right.\\
    & \qquad \left. - \bSigma(\rho)^{-1}\bGamma\bSigma(\rho)^{-1} \rp\Lp \text{d}\bSigma(\rho) \Rp\bSigma(\rho)^{-1}\widehat{\bx}(\btheta)\\
    \frac{\partial^2 \text{pl}}{\partial \rho \partial \lambda_e^2} &= \lambda^2\widehat{\bx}(\btheta)^{\sf T}\bSigma(\rho)^{-1}\Lp \text{d}\bSigma(\rho) \Rp\bSigma(\rho)^{-1}\bGamma\bA^{\sf T}\widehat{\bpsi}(\btheta)\\
    \frac{\partial^2 \text{pl}}{\partial \lambda_e^4} &= -\frac{p}{\lambda_e^4} + \widehat{\bpsi}(\btheta)^{\sf T}\bA\bGamma\bA^{\sf T}\widehat{\bpsi}(\btheta),
\end{split}
\eeq where $\text{d}^2\text{L}$ represents the second derivative of $\logdet \bSigma(\rho)$ with respect to $\rho$ and $\text{d}^2\bSigma(\rho)$ is the second derivative of $\bSigma(\rho)$ with respect to $\rho$. $\text{d}^2\bSigma(\rho)$ also has a BTTB structure as $\bSigma(\rho)$ and $\text{d}\bSigma(\rho)$.

These entries are then approximated using the approximation to $\bGamma$ as presented in \cite{chung2018efficient}, namely \beq \label{eq: bgamma_approx} \bGamma \approx \lambda^{-2}\Lp \bSigma(\rho) - \bZ_k\bDelta_k\bZ_k^{\sf T} \Rp, \eeq where $\bZ_k = \bSigma(\rho)\bV_k\bW_k$ with $\bB_k^{\sf T}\bB_k = \bW_k\bTheta_k\bW_k$ and $\bDelta_k = \Lp \bI + \lambda^{-2}\bTheta_k \Rp^{-1}$.

We define $\bz_0 = \by - \bX\bbeta - \bA\bx^*_k(\btheta) = \bpsi^*_k(\btheta)$. The approximated entries of the Hessian are \beq \label{eq: Hess_approx}
\begin{split}
    \frac{\partial^2 \text{pl}}{\partial \bbeta \partial \bbeta^{\sf T}} &\approx -\lambda_e^2\bX^{\sf T}\bX + \frac{\lambda_e^4}{\lambda^2}\bX^{\sf T}\bA\bSigma(\rho)\bA^{\sf T}\bX\\
    & \quad - \frac{\lambda_e^4}{\lambda^2}\bX^{\sf T}\bU_k\bB_k\bW_k\bDelta_k\bW_k^{\sf T}\bB_k^{\sf T}\bU_k^{\sf T}\bX\\
    \frac{\partial^2 \text{pl}}{\partial \bbeta \partial \lambda^2} &\approx \frac{\lambda_e^2}{\lambda^2}\bX^{\sf T}\bA\widehat{x}^*(\btheta) - \frac{\lambda_e^2}{\lambda^2}\bX^{\sf T}\bU_k\bB_k\bW_k\bDelta_k\bW_k^{\sf T}\bz_k\\
    \frac{\partial^2 \text{pl}}{\partial \bbeta \partial \rho} &\approx -\lambda_e^2\bX^{\sf T}\bA\Lp \text{d}\bSigma(\rho) \Rp\bV_k\bz_k\\
    & \quad + \lambda_e^2\bX^{\sf T}\bU_k\bB_k\bW_k\bDelta_k\bW_k^{\sf T}\bV_k^{\sf T}\Lp \text{d}\bSigma(\rho) \Rp\bV_k\bz_k\\
    \frac{\partial^2 \text{pl}}{\partial \bbeta \partial \lambda_e^2} &\approx -\bX^{\sf T}\bz_0 - \frac{\lambda_e^2}{\lambda^2}\bX^{\sf T}\bA\bSigma(\rho)\bA^{\sf T}\bz_0\\
    & \quad + \frac{\lambda_e^2}{\lambda^2}\bX^{\sf T}\bU_k\bB_k\bW_k\bDelta_k\bW_k^{\sf T}\bB_k^{\sf T}\bU_k^{\sf T}\bz_0\\
    \frac{\partial^2 \text{pl}}{\partial \lambda^4} &\approx -\frac{n}{2\lambda^4} + \frac{1}{\lambda^2}\|\bz_k\|_2^2 - \frac{1}{\lambda^2}\bz_k^{\sf T}\bW_k\bDelta_k\bW_k^{\sf T}\bz_k\\
    \frac{\partial^2 \text{pl}}{\partial \lambda^2 \partial \rho} &\approx -\frac{1}{2}\bz_k^{\sf T}\bV_k^{\sf T}\Lp \text{d}\bSigma(\rho) \Rp\bV_k\bz_k\\
    & \quad + \bz_k^{\sf T}\bV_k^{\sf T}\Lp \text{d}\bSigma(\rho) \Rp\bV_k\bW_k\bDelta_k\bW_k^{\sf T}\bz_k\\
    \frac{\partial^2 \text{pl}}{\partial \lambda^2 \partial \lambda_e^2} &\approx -\frac{1}{\lambda^2}\bz_k^{\sf T}\bB_k^{\sf T}\bU_k^{\sf T}\bz_0 + \frac{1}{\lambda^2}\bz_k^{\sf T}\bW_k^{\sf T}\bDelta_k\bW_k^{\sf T}\bB_k^{\sf T}\bU_k^{\sf T}\bz_0\\
    \frac{\partial^2 \text{pl}}{\partial \rho^2} &\approx -\frac{1}{2}\widehat{\text{d}^2\text{L}} + \frac{\lambda^2}{2}\bz_k^{\sf T}\bV_k^{\sf T}\Lp \text{d}^2\bSigma(\rho) \Rp\bV_k\bz_k\\
    & \quad - \lambda^2 \bz_k^{\sf T}\bV_k^{\sf T}\Lp \text{d}\bSigma(\rho) \Rp\bV_k\bW_k\bDelta_k\bW_k^{\sf T}\bV_k^{\sf T}\Lp \text{d}\bSigma(\rho) \Rp\bV_k\bz_k\\
    \frac{\partial^2 \text{pl}}{\partial \rho \partial \lambda_e^2} &\approx \bz_k^{\sf T}\bV_k^{\sf T}\Lp \text{d}\bSigma(\rho) \Rp\bA^{\sf T}\bz_0\\
    & \quad - \bz_k^{\sf T}\bV_k^{\sf T}\Lp \text{d}\bSigma(\rho) \Rp\bV_k\bW_k\bDelta_k\bW_k^{\sf T}\bB_k^{\sf T}\bU_k^{\sf T}\bz_0\\
    \frac{\partial^2 \text{pl}}{\partial \lambda_e^4} &\approx -\frac{p}{2\lambda_e^4} + \frac{1}{\lambda^2}\bz_0^{\sf T}\bA\bSigma(\rho)\bA^{\sf T}\bz_0\\
    & \quad - \frac{1}{\lambda^2}\bz_0^{\sf T}\bU_k\bB_k\bW_k\bDelta_k\bW_k^{\sf T}\bB_k^{\sf T}\bU_k^{\sf T}\bz_0,
\end{split}
\eeq where $\widehat{\text{d}^2\text{L}}$ is a numerical approximation to $\text{d}^2\text{L}$. We do not use this approximation for our computing, but hope to use it in future.

\section{Additional Tables from the Simulation Study}
\label{app:addl_tab}

In this section, we provide additional results for the simulation study. Table \ref{tab: rmse_sim1} evaluates parameter estimations for the first simulation study for both SPDE and Kryging methods. The same is done in Tables \ref{tab: rmse_sim2} and \ref{tab: rmse_sim3} for the second and third simulation studies. The results across the board are similar as mentioned in Section \ref{sec:sim}. SPDE performs better in estimating the nugget parameter $\tau^2$, while Kryging performs better in estimating the partial sill parameter $\sigma^2$. Both methods do equally well in estimating the mean parameter $\beta$ and the spatial range parameter $\rho$.

\begin{table}[ht]
\centering
\caption{RMSE in estimating the parameters for SPDE and Kryging for different gridsizes and choices of $k$ as in the first simulation study. The true values for the parameters were $(44.49, 3, 0.5, 1)$. The figures in brackets indicate standard error.}
\label{tab: rmse_sim1}
\begin{adjustbox}{max width=\linewidth}
\begin{tabular}{c|c|ccccc}
  \hline
 Parameter & Grid Size & SPDE & \multicolumn{4}{c}{Kryging}\\
  \cline{4-7}
 & & & k=20 & k=50 & k=100 & k=200\\ 
 \hline
$\beta$ & $100 \times 100$ & 0.30 (0.30) & 0.31 (0.32) & 0.30 (0.32) & 0.31 (0.32) & 0.31 (0.32) \\ 
  & $200 \times 200$ & 0.23 (0.22) & 0.28 (0.23) & 0.28 (0.23) & 0.28 (0.23) & 0.28 (0.23) \\ 
  & $300 \times 300$ & 0.32 (0.24) & 0.32 (0.25) & 0.32 (0.25) & 0.32 (0.25) & 0.32 (0.25) \\ 
  & $400 \times 400$ & 0.26 (0.26) & 0.29 (0.26) & 0.29 (0.26) & 0.29 (0.26) & 0.29 (0.26) \\
 \hline
$\sigma^2$ & $100 \times 100$ & 1.43 (0.11) & 0.36 (0.34) & 0.36 (0.34) & 0.36 (0.34) & 0.36 (0.34) \\ 
  & $200 \times 200$ & 1.59 (0.07) & 0.31 (0.24) & 0.31 (0.24) & 0.31 (0.24) & 0.31 (0.24) \\ 
  & $300 \times 300$ & 1.70 (0.07) & 0.33 (0.26) & 0.33 (0.26) & 0.33 (0.26) & 0.33 (0.26) \\ 
  & $400 \times 400$ & 1.81 (0.06) & 0.30 (0.17) & 0.30 (0.17) & 0.30 (0.17) & 0.30 (0.17) \\ 
  \hline
 $\tau^2$& $100 \times 100$ & 0.10 (0.01) & 0.17 (0.05) & 0.17 (0.05) & 0.17 (0.05) & 0.17 (0.05) \\ 
 & $200 \times 200$ & 0.06 (0.01) & 0.16 (0.04) & 0.16 (0.04) & 0.16 (0.04) & 0.16 (0.04) \\ 
 & $300 \times 300$ & 0.05 (0.00) & 0.16 (0.05) & 0.16 (0.05) & 0.16 (0.05) & 0.16 (0.05) \\ 
 & $400 \times 400$ & 0.04 (0.00) & 0.19 (0.03) & 0.19 (0.03) & 0.19 (0.03) & 0.19 (0.03) \\ 
     \hline
$\rho$ & $100 \times 100$ & 0.06 (0.02) & 0.03 (0.00) & 0.03 (0.00) & 0.02 (0.00) & 0.02 (0.00) \\ 
  & $200 \times 200$ & 0.01 (0.01) & 0.03 (0.00) & 0.03 (0.00) & 0.03 (0.00) & 0.03 (0.00) \\ 
  & $300 \times 300$ & 0.01 (0.01) & 0.03 (0.00) & 0.03 (0.00) & 0.03 (0.00) & 0.03 (0.00) \\ 
  & $400 \times 400$ & 0.03 (0.00) & 0.03 (0.00) & 0.03 (0.00) & 0.03 (0.00) & 0.03 (0.00) \\ 
   \hline
\end{tabular}\end{adjustbox}
\end{table}

\begin{table}
\centering
\caption{RMSE in estimating the parameters for SPDE and Kryging under different parametric settings and different choices of $k$ as in the second simulation study. The true values for the parameters were $(44.49,3,0.5,0.05)$, $(44.49,3,0.5,0.2)$, $(44.49,1.5,0.5,0.1)$ and $(44.49,6,0.5,0.1)$ for settings 1 through 4 respectively. The figures in brackets indicate standard error.}
\label{tab: rmse_sim2}
\begin{adjustbox}{max width=\linewidth}
\begin{tabular}{c|c|ccccc}
  \hline
 Parameter & Setting & SPDE & \multicolumn{4}{c}{Kryging}\\
  \cline{4-7}
 & & & k=20 & k=50 & k=100 & k=200\\
 \hline
$\beta$ & Setting 1 & 0.16 (0.15) & 0.16 (0.14) & 0.16 (0.14) & 0.16 (0.14) & 0.16 (0.14) \\ 
  & Setting 2 & 0.42 (0.37) & 0.46 (0.39) & 0.46 (0.39) & 0.46 (0.39) & 0.46 (0.39) \\ 
  & Setting 3 & 0.18 (0.10) & 0.22 (0.10) & 0.22 (0.10) & 0.22 (0.10) & 0.22 (0.10) \\ 
  & Setting 4 & 0.42 (0.27) & 0.43 (0.28) & 0.43 (0.28) & 0.43 (0.28) & 0.43 (0.28) \\
 \hline
$\sigma^2$ & Setting 1 & 1.43 (0.05) & 0.16 (0.17) & 0.16 (0.17) & 0.16 (0.17) & 0.16 (0.17) \\ 
  & Setting 2 & 1.79 (0.11) & 0.59 (0.32) & 0.59 (0.32) & 0.59 (0.32) & 0.59 (0.32) \\ 
  & Setting 3 & 0.47 (0.06) & 0.23 (0.21) & 0.23 (0.21) & 0.23 (0.21) & 0.23 (0.21) \\ 
  & Setting 4 & 4.07 (0.06) & 0.70 (0.44) & 0.70 (0.44) & 0.70 (0.44) & 0.70 (0.44) \\
 \hline
$\tau^2$ & Setting 1 & 0.10 (0.01) & 0.16 (0.02) & 0.16 (0.02) & 0.16 (0.02) & 0.16 (0.02) \\ 
  & Setting 2 & 0.04 (0.01) & 0.21 (0.06) & 0.21 (0.06) & 0.21 (0.06) & 0.21 (0.06) \\ 
  & Setting 3 & 0.07 (0.15) & 0.30 (0.03) & 0.30 (0.03) & 0.30 (0.03) & 0.30 (0.03) \\ 
  & Setting 4 & 0.12 (0.01) & 0.10 (0.06) & 0.10 (0.06) & 0.10 (0.06) & 0.10 (0.06) \\ 
   \hline
   $\rho$ & Setting 1 & 0.03 (0.00) & 0.02 (0.00) & 0.02 (0.00) & 0.02 (0.00) & 0.02 (0.00) \\ 
  & Setting 2 & 0.05 (0.02) & 0.13 (0.00) & 0.13 (0.00) & 0.13 (0.00) & 0.13 (0.00) \\ 
  & Setting 3 & 0.03 (0.02) & 0.03 (0.00) & 0.03 (0.00) & 0.03 (0.00) & 0.03 (0.00) \\ 
  & Setting 4 & 0.01 (0.00) & 0.03 (0.00) & 0.03 (0.00) & 0.03 (0.00) & 0.03 (0.00) \\ 
   \hline
\end{tabular}\end{adjustbox}
\end{table}

\begin{table}
\centering
\caption{RMSE in estimating the parameters for SPDE and Kryging for different choices of underlying gridsize and $k$ for the simulation study with irregularly spaced data. The true parameter values were $(44.49, 3, 0.5, 0.1)$. The figures in brackets indicate standard error.}
\label{tab: rmse_sim3}
\begin{adjustbox}{max width=\linewidth}
\begin{tabular}{c|c|cccc}
\hline
 Parameter & Gridsize & SPDE & \multicolumn{3}{c}{Kryging}\\
  \cline{4-6}
 & &  & k=20 & k=50 & k=100\\
 \hline
 $\beta$ & $200 \times 200$ & 0.24 (0.18) & 0.29 (0.16) & 0.29 (0.16) & 0.29 (0.16) \\ 
 & $300 \times 300$ & 0.24 (0.18) & 0.29 (0.16) & 0.29 (0.16) & 0.29 (0.16) \\ 
 & $400 \times 400$ & 0.24 (0.18) & 0.29 (0.16) & 0.29 (0.16) & 0.29 (0.16) \\ 
  \hline
 $\sigma^2$ & $200 \times 200$ & 1.61 (0.07) & 0.26 (0.21) & 0.26 (0.21) & 0.26 (0.21) \\ 
 & $300 \times 300$ & 1.61 (0.07) & 0.26 (0.21) & 0.26 (0.21) & 0.26 (0.21) \\ 
 & $400 \times 400$ & 1.61 (0.07) & 0.26 (0.21) & 0.26 (0.21) & 0.26 (0.21) \\
  \hline
 $\tau^2$ & $200 \times 200$ & 0.06 (0.01) & 0.17 (0.04) & 0.17 (0.04) & 0.17 (0.04) \\ 
 & $300 \times 300$ & 0.06 (0.01) & 0.17 (0.04) & 0.17 (0.04) & 0.17 (0.04) \\ 
 & $400 \times 400$ & 0.06 (0.01) & 0.17 (0.04) & 0.17 (0.04) & 0.17 (0.04) \\ 
  \hline
  $\rho$ & $200 \times 200$ & 0.01 (0.01) & 0.03 (0.00) & 0.03 (0.00) & 0.03 (0.00) \\ 
 & $300 \times 300$ & 0.01 (0.01) & 0.03 (0.00) & 0.03 (0.00) & 0.03 (0.00) \\ 
 & $400 \times 400$ & 0.01 (0.01) & 0.03 (0.00) & 0.03 (0.00) & 0.03 (0.00) \\ 
  \hline
\end{tabular}\end{adjustbox}
\end{table}

\end{document}